\documentclass[journal=jpcbfk,manuscript=article]{achemso}

\usepackage[version=3]{mhchem} 
\usepackage[]{bm}
\usepackage{amssymb}

\usepackage{xr}
\usepackage{hyperref}
\usepackage{cleveref}
\usepackage{siunitx}
\usepackage{upgreek}
\usepackage{array}
\usepackage{tabularx}
\usepackage{xcolor}

\usepackage[textsize=tiny]{todonotes}

\DeclareSIUnit\angstrom{\text {Å}}

\usepackage[font=scriptsize]{caption}

\SectionNumbersOn

\Crefname{figure}{Figure}{Figures}

\Crefname{equation}{eq}{eqs}

\author{Wesley W. Oliver}
\affiliation[Princeton University]
{Department of Chemical and Biological Engineering, Princeton University, Princeton, NJ 08544, USA}
\author{William M. Jacobs}
\affiliation[Princeton University]
{Department of Chemistry, Princeton University, Princeton, NJ 08544, USA}
\author{Michael A. Webb}
\email{mawebb@princeton.edu}
\affiliation[Princeton University]
{Department of Chemical and Biological Engineering, Princeton University, Princeton, NJ 08544, USA}

\title{When $B_2$ is Not Enough: Evaluating Simple Metrics for Predicting Phase Separation of Intrinsically Disordered Proteins}

\begin{document}

\begin{abstract}
    Understanding and predicting the phase behavior of intrinsically disordered proteins (IDPs) is of significant interest due to their role in many biological processes. 
    However, effectively characterizing phase behavior and its complex dependence on protein primary sequence remains challenging. 
    In this study, we evaluate the efficacy of several simple computational metrics to quantify the propensity of single-component IDP solutions to phase separate; 
    specific metrics considered include the single-chain radius of gyration, the second virial coefficient, and a newly proposed quantity termed the expenditure density.
    Each metric is computed using coarse-grained molecular dynamics simulations for 2,034 IDP sequences. 
    Using machine learning, we analyze this data to understand how sequence features correlate with the predictive performance of each metric and to develop insight into their respective strengths and limitations. 
    The expenditure density is determined to be a broadly useful metric that combines simplicity, low computational cost, and accuracy; it also provides a continuous measure that remains informative across both phase-separating and non–phase-separating sequences.
    Additionally, this metric shows promise in its ability to improve predictions of other properties for IDP systems.
    This work extends existing literature by advancing beyond binary classification, which can be useful for rapidly screening phase behavior or predicting other properties of IDP-related systems.
\end{abstract}

\section{Introduction\label{sec:intro}}

Cells achieve spatiotemporal organization of thousands of biomolecules through the dynamic and reversible formation of biomolecular condensates, which compartmentalize and concentrate specific cellular cargo without requiring lipid membranes~\cite{banani_biomolecular_2017}.
Many condensates, such as nucleoli~\cite{feric_coexisting_2016} and P-granules~\cite{brangwynne_germline_2009}, form, at least in part, by spontaneous phase separation into biopolymer-rich and biopolymer-lean phases~\cite{mittag_conceptual_2022}; this is often associated with liquid-liquid phase separation.
Condensate formation and regulation are critical for cellular function, whereas dysregulation is implicated in neurodegenerative diseases, infectious diseases, and cancers~\cite{babinchak_role_2019,alberti_liquidliquid_2019}.
Therefore, understanding and controlling the formation of condensates could inform possible therapeutic strategies or otherwise facilitate the design of synthetic condensates with bespoke functionalities~\cite{do_engineering_2022,qian_synthetic_2022}. 

Predicting the phase behavior of intrinsically disordered proteins (IDPs) and regions (IDRs) from sequence remains a major challenge, despite growing interest in understanding their role in biomolecular condensates. The absence of stable structure in IDPs enables dynamic, multivalent interactions that can drive liquid–liquid phase separation~\cite{oldfield_intrinsically_2014,holehouse_molecular_2024,pappu_phase_2023}. Numerous studies have explored how composition, mutations, temperature, and ion concentration affect phase behavior in well-known systems such as DDX4, FUS, and LAF-1~\cite{nott_phase_2015,wang_molecular_2018,schuster_identifying_2020}. Complementary insights have come from theoretical models grounded in polymer physics~\cite{brangwynne_polymer_2015,lin_phase_2017} and from minimal model systems that reveal how specific sequence features promote or inhibit condensation~\cite{statt_model_2020,rekhi_role_2023,das_conformations_2013}. More recently, coarse-grained molecular simulations using models such as HPS-Urry~\cite{regy_improved_2021}, Mpipi~\cite{joseph_physics-driven_2022}, and CALVADOS~\cite{tesei_accurate_2021,tesei_improved_2022} have enabled quantitative comparisons with experimental data. Despite these advances, accurately and efficiently predicting phase behavior for arbitrary IDP sequences remains an unresolved problem.

Machine learning and data-driven approaches are increasingly used to complement theory, simulation, and experiments in predicting and understanding phase behavior.
Machine learning models trained on curated databases, such as LLPSDB~\cite{li_llpsdb_2020} and PhaSePro~\cite{meszaros_phasepro_2020}, have produced various tools that predict IDP phase behavior directly from primary sequence (e.g.,  PSAP~\cite{van_mierlo_predicting_2021}, PSPredictor~\cite{chu_prediction_2022}, PhaSePred~\cite{chen_screening_2022}, and PSPHunter~\cite{sun_precise_2024}). 
Retrospective analysis of model predictions has helped to elucidate sequence characteristics that contribute to phase separation~\cite{chen_screening_2022,sun_precise_2024}.
However, the paucity of data as well as selection bias, particularly for phase-separating IDPs in experimental settings, calls into question the generalizability of these models and insights.
Consequently, high-throughput molecular simulation, often coupled with active learning, has emerged as a strategy to augment training data and thus characterize the properties of a broader distribution of IDPs~\cite{tesei_conformational_2024,an_active_2024,von_bulow_prediction_2025,changiarath_sequence_2025,lotthammer_direct_2024}.
This data-driven strategy presents a general opportunity to examine the phase behavior not only of naturally occurring IDPs but also of sequences absent from known biological genomes~\cite{pesce_design_2024, an_active_2024}, potentially enabling a more comprehensive understanding of the sequence determinants of phase behavior.

The concept of a ``propensity'' for phase separation is central to many studies regarding biological condensates. 
While the idea of phase-separation propensity is qualitatively intuitive, as some proteins phase separate while others do not under the same conditions, there is no consensus regarding  a reliable quantitative measure.
Given two IDPs that both undergo phase separation, does one phase separate more ``strongly''?
Similarly, if neither IDP phase separates, can one be considered ``closer'' to the phase-separating regime?
Ideally, metrics for phase-separation propensity should enable classification of phase behavior and also provide continuous information on the degree of phase-separation or lack thereof.
The critical temperature ($T_\mathrm{c}$) provides one such quantitative measure of phase-separation propensity; however, its calculation is non-trivial and computationally laborious,~\cite{dignon_sequence_2018} which limits its utility for data-driven investigations.
Alternatively, various proxies for $T_\mathrm{c}$ have been used, including the single-chain radius of gyration ($R_\mathrm{g}$)~\cite{lin_phase_2017,zeng_connecting_2020},  theta temperature ($T_\mathrm{\theta}$)~\cite{dignon_relation_2018,zeng_connecting_2020,zeng_design_2021}, Flory scaling exponent ($\nu$)~\cite{tesei_conformational_2024}, second virial coefficient ($B_2$)~\cite{quigley_second_2015,rekhi_role_2023}, Boyle temperature ($T_\mathrm{B}$)~\cite{dignon_relation_2018,adachi_predicting_2024}, and condensed phase-dilute phase transfer free energy ($\Delta g$)~\cite{von_bulow_prediction_2025,changiarath_sequence_2025}. 
While these metrics are computationally more tractable than $T_\mathrm{c}$, there are notable breakdowns in terms of their discriminatory power with respect to correlating with phase behavior.\cite{von_bulow_prediction_2025,dignon_temperature-controlled_2019,pal_differential_2024,rekhi_role_2023}
For example, many sequences may possess equivalent $B_2$ but exhibit disparate phase behavior outcomes.\cite{jin_predicting_2025}
This limits their utility as features in the training of machine-learning models for predicting phase behavior.
Assessing the performance of existing metrics and exploring new alternatives across a diverse distribution of IDPs may help improve quantitative assessments of phase behavior.

In this work, we systematically evaluate the efficacy of several simple metrics to predict the propensity of phase separation for single-component IDP solutions.
This assessment leverages data generated from coarse-grained molecular dynamics simulations of 2,034 polypeptide sequences~\cite{webb_thermodynamic_2023} and compares three metrics: the single-chain radius of gyration, the second virial coefficient, and a proposed metric referred to as the expenditure density.
We compare the utility of metrics across three tasks.
First, we assess how accurately the metric by itself, assumed to be known without error, predicts whether a given sequence undergoes phase separation in solution. 
Second, we determine how well the metrics can be predicted from intrinsic sequence features using neural-network surrogate models; this strategy would be warranted in high-throughput assessment scenarios.
Analysis of these models enables us to identify which features are most influential for predicting metric values and how these features compare to those most influential for predicting phase separation directly.
Third, we examine whether the prediction of other IDP properties can be improved by including metrics as additional features supplied to surrogate models.
This is  specifically tested for the case of chain self-diffusivity in the condensed phase. 
Finally, we examine specific failure modes and their sequence dependence to reveal the strengths and limitations of each metric. 
This analysis offers a foundation for more accurate, data-efficient prediction and screening of phase behavior in IDP solutions and related systems. 

\section{Methods\label{sec:methods}}
\subsection{Summary of IDP sequences\label{subsec:seqs}}
We analyzed 2034 IDP sequences ranging in length from 20 to 50 residues~\cite{webb_thermodynamic_2023}. 
Of these, 1,266 are naturally occurring sequences from the DISPROT database~\cite{sickmeier_disprot_2007}; the remaining 768 were generated during an active learning campaign that explored the relationship between condensed-phase diffusivity and the second virial coefficient~\cite{an_active_2024}. 
This latter abiological set of sequences spans a range of sequence characteristics, from highly charged polyampholytes to neutral hydrophobic sequences. 
In addition to including both phase-separating and non-phase-separating sequences, the condensed phases are also known span a range of viscosities by intention of the active learning campaign.
We note that neither the DISPROT database~\cite{sickmeier_disprot_2007} nor the active learning-derived dataset~\cite{webb_thermodynamic_2023} was intentionally crafted to provide balanced coverage of sequence diversity. 
Nevertheless, the collection of sequences across both sources enables evaluation of phase-separation propensity across IDPs with varied sequence lengths, composition, patterning, and properties.

\subsection{Simulation Details\label{subsec:sim_details}}

\paragraph{Model of protein sequences.}chains
All simulations employed the coarse-grained HPS-Urry model,~\cite{regy_improved_2021} which describes residue-residue interactions within an implicit solvent framework.
The force field consists of a harmonic bonded potential as well as short-range van der Waals (vdW) and long-range electrostatic (el) nonbonded interactions,
\begin{equation} \label{eqn:1}
  U_{\text{tot}} = \sum_i k_{\text{b}}(r_{i,i+1}-b_0)^2+\sum_{i,j}\phi^{\text{vdW}}(r_{ij})+\sum_{i,j}\phi^{\text{el}}(r_{ij}),
\end{equation}
where $r_{i,i+1}$ is the distance between bonded residues $i$ and $i+1$, the force constant is $k_{\text{b}}=\SI{10}{kcal/(mol\cdot \AA^2)}$, and the equilibrium bond length is $b_0=3.82$ \AA.
Short-range vdW interactions are described by the Ashbaugh-Hatch potential:
\begin{equation}  \label{eqn:2}
  \phi^{\text{vdW}}(r_{ij})=\begin{cases}
  \phi^{\text{LJ}}(r_{ij})+(1-\lambda_{ij})\epsilon &\quad r_{ij} \leq 2^{1/6}\sigma_{ij}\\
  \lambda_{ij}\phi^{\text{LJ}}(r_{ij}) &\quad  r_{ij} > 2^{1/6}\sigma_{ij},
  \end{cases}
\end{equation}
where $r_{ij}$ is the distance between nonbonded residues $i$ and $j$, and $\phi^{\text{LJ}}(r_{ij}) = 4\varepsilon[(\sigma_{ij}/r_{ij})^{12} - (\sigma_{ij}/r_{ij})^6]$ is the Lennard-Jones potential with $\varepsilon=0.2$ kcal/mol.
The interaction strengths, $\lambda_{ij}$, are derived from the Urry hydrophobicity scale~\cite{urry_hydrophobicity_1992} with additional optimizations described by Regy et al.~\cite{regy_improved_2021}
The distance parameters, $\sigma_{ij}$, 
are determined from an arithmetic mixing rule of the prescribed vdW diameters.  
The screened electrostatic interactions take the form
\begin{equation} \label{eqn:4}
  \phi^{\text{el}}(r_{ij}) = \frac{q_i q_j}{4\pi \epsilon_\text{r} r_{ij}} e^{-\kappa r_{ij}},
\end{equation}
where $q_i$ is the charge of residue $i$.
In this study, the dielectric constant was set as $\epsilon_\text{r}=80$, and the Debye screening length was set as $\kappa=10$ \text{\AA} to reflect an ion concentration of 100 mM in water. 

\hfill

\paragraph{General simulation protocols.}
All simulations were run using the October 29, 2020 release of LAMMPS~\cite{thompson_lammps_2022}.
Equations of motion were evolved using the velocity-Verlet algorithm with a 10 \unit{\femto\second} time step.
Simulations were performed in the isochoric-isothermal (NVT) ensemble at a temperature of 300 K.
The temperature was regulated using a Langevin thermostat with a damping factor of 1000 \unit{\femto\second}.

\hfill

\paragraph{Assessment of phase behavior.}
Rigorous assessment of phase behavior in IDP systems typically relies on direct-coexistence simulations using slab geometries across multiple conditions. Due to the high computational cost of this approach, we instead evaluated phase behavior by constructing an approximate equation of state (EoS) from simulations at varying densities and identifying the presence of a negative pressure loop.~\cite{binder_beyond_2012}. 
A negative pressure loop indicates that the system likely phase separates at those conditions, while its absence suggests existence as a single, homogeneous phase.
This approach also enables inference of an expected condensed-phase density, $\rho_\text{c}$, based on the highest density at which zero pressure is observed (assuming the vapor pressure is low). 
This approach has demonstrated good agreement with results from more expensive direct-coexistence simulations~\cite{an_active_2024}.

To obtain the EoS, we used the following procedure.
First, systems were initialized by successively placing 100 identical chains with random orientations in a cubic, periodic box with side lengths adjusted to obtain densities beginning at $\rho=0.05$ \unit{\gram\per\mL} and then ranging from $\rho=0.1$ \unit{\gram\per\mL} up to $\rho=1.4$ \unit{\gram\per\mL} in increments of $0.1$ \unit{\gram\per\mL}. 
Chain placements were accepted if all monomers were at least 1 \unit{\angstrom} away from all other monomers.
If chain placement was unsuccessful after 99 attempts, a random placement was assigned and followed by a minimization using a soft-core potential until all monomers were separated by at least 2 \unit{\angstrom}.
The starting chain geometry was constructed by consecutively placing residues at a distance of 3.82 \unit{\angstrom} and a random angle of up to 0.3 \unit{\pi} radians.
Second, systems were subjected to 5000 steps of energy minimization via conjugate gradient descent.
Third, simulations in the NVT ensemble were performed for 100 \unit{\nano\second}.
The first half of each trajectory was discarded as equilibration, and the latter half was used for production data.
For production data, the instantaneous stress tensor was sampled every 1 \unit{\pico\second} to compute an average pressure $P(\rho) = \langle \textbf{P}_{ii}\rangle_\rho$, where $i$ indexes one of the $x$, $y$, or $z$ components.
Standard errors for $P$ were computed by splitting the production run into five blocks and treating the average of each block as an independent sample.

The presence of a negative pressure loop was determined by observing statistically significant negative pressure values, defined by a one-sided t-test with a p-value less than 0.01.
Example $P(\rho)$ curves for sequences with different phase behavior are shown in the Supporting Information, Figure S1. 
The EoS was represented as a continuous function by constructing a cubic spline between all $\left(\rho,P\right)$ pairs and an additional pair $\left(0,0\right)$ using the CubicSpline object from the SciPy Python package~\cite{virtanen_scipy_2020}. 
The boundary conditions of the cubic spline were fixed such that the first derivative at $\left(0,0\right)$ is zero.
If a sequence was assessed as phase-separating, $\rho_\mathrm{c}$ was determined as the largest density satisfying $P(\rho)$ = 0. 
Standard errors for $\rho_\mathrm{c}$ of phase-separating sequences were calculated by bootstrapping $10^4$ additional EoS calculations from the production pressure values.

\subsection{Metrics of Phase-Separation Propensity\label{subsec:propensity_metrics}}

\paragraph{Radius of gyration, $R_\text{g}$.}
The radius of gyration of a polymer chain characterizes its overall size based on the average distance of monomers from the chain center-of-mass.
There is often a presumed connection between the conformational state of a polymer chain and its macroscopic phase behavior.\cite{R:2014_Wang_Theory,R:2021_Xu_Coil,R:2024_Dhamankar_Asymmetry}
For a given sequence length $N$, $R_\mathrm{g}$ reflects the degree of intramolecular self-interaction. To create a normalized, intensive measure, we compare $R_\mathrm{g}$ to that of an ideal chain, defined as $R_\text{g}^{\text{id}} \equiv \sqrt{\langle R_\text{g}^2 \rangle_\text{id}} = b_0\sqrt{N/6}$.\cite{rubinstein_polymer_2003} We then define the normalized radius of gyration as $\tilde{R}_\mathrm{g} = R_\mathrm{g} / R_\text{g}^{\text{id}}$.
A related measure is the Flory scaling exponent $\nu$, since $\log \tilde{R}_\text{g} \propto (\nu - 1/2) \log N$, where the prefactor expresses a deviation from theta conditions.
Consequently, $\tilde{R}_\text{g} < 1$ typically corresponds to poor-solvent conditions and increased phase separation propensity, while $\tilde{R}_\text{g} > 1$ suggests good-solvent conditions and reduced phase-separation propensity.\cite{raos_chain_1996,rubinstein_polymer_2003,dignon_relation_2018}
We find that $\tilde{R}_\mathrm{g}$ outperforms $R_\mathrm{g}$ in characterizing phase behavior (Supporting Information, Figure S2) and thus adopt it for further analysis.

For any given sequence, $R_\mathrm{g}$ was calculated by placing a single chain in a cubic, periodic box with side lengths set such that the density was equal to $\rho=1\times10^{-5}$ \unit{\gram\per\mL}. 
The simulation was equilibrated for 100 ns followed by a 900 ns production run; coordinates were recorded every 50 \unit{\pico\second}.
Given sets of residue coordinates $\{\vec{r}_i\}$ and masses $\{m_i\}$, the instantaneous radius of gyration $R_\mathrm{g}(t)$ was calculated as
\begin{align}
R_\mathrm{g}(t) &= \sqrt{\frac{\sum_{i=1}^{N} m_i |\vec{r}_i - \vec{r}_{\text{cm}}|^2}{\sum_{i=1}^{N} m_i}} \\
\vec{r}_{\text{cm}} &= \frac{\sum_{i=1}^{N} m_i \vec{r}_i}{\sum_{i=1}^{N} m_i}. 
\end{align}
Values reported herein correspond to $\langle R_\mathrm{g}(t) \rangle$, which was determined using all samples collected during production.
Standard errors were computed by splitting the production run into five blocks and treating the average of each block as an independent sample.

\hfill

\paragraph{Second virial coefficient, $B_2$.}
The second virial coefficient quantifies the net interaction between two polymer chains in dilute solution.
In general, negative values indicate pairwise attraction, while positive values indicate pairwise repulsion. 
As a metric for phase-separation propensity, more negative $B_2$ should correlate with a higher likelihood of phase separation; however, even substantially negative $B_2$ does not guarantee phase separation\cite{rekhi_role_2023,an_active_2024,jin_predicting_2025}. 
For single-component systems, $B_2 > 0$ would make phase separation generally unlikely.

For any given sequence, $B_2$ can be calculated via
\begin{equation}\label{eq:b2}
    B_2=2\pi\int_0^\infty dr \,\,r^2\left[1-e^{\beta w(r)}\right]
\end{equation}
where $w(r)$ is the potential of mean force between two chains as a function of $r$, the distance between the center-of-mass of each chain.~\cite{rubinstein_polymer_2003}.
Here, $w(r)$ was obtained using the adaptive biasing force (ABF) method~\cite{comer_adaptive_2015}, as implemented in the COLVARS package in LAMMPS~\cite{thompson_lammps_2022, fiorin_using_2013}. Systems were initialized with two identical chains in a periodic cubic box (300 \unit{\angstrom} per side). Thirty independent ABF trajectories, each with 5 \unit{\micro\second} of production data, were run to estimate standard errors.
Biasing forces were applied over distances from $r = 1$ to $100$ \unit{\angstrom}; this distance range was discretized into 1 \unit{\angstrom} bins for the purpose of gathering statistics.
Biasing forces at a given distance were scaled by a factor  $\alpha(r) = \min\!\biggl(\frac{n(r)}{n_\text{min}},\,1\biggr)$
where \(n(r)\) is the number of samples collected for the bin at distance $r$ and $n_\text{min}$ is minimum threshold number of samples required for the biasing forces to act at full strength;
we use $n_\text{min} = 1000$.
For numerical integration of eq. \eqref{eq:b2}, $w(r)$ was set to zero for $r > 60$ \unit{\angstrom} due to limited sampling at large $r$.

Because $B_2$ is an extensive quantity that is expected to scale with sequence length, we define $\tilde{B}_2=B_2/V_0$ as an intensive quantity to capture intrinsic behavior.
Here, the normalization stems from the pervaded volume of an ideal polymer chain, $V_0=\left(4\pi/3\right)b_0^3\left(N/6\right)^\frac{3}{2}$~\cite{rubinstein_polymer_2003}, where $b_0$ is the equilibrium bond length and $N$ is the number of monomers.
For the purpose of this calculation, we approximated $b_0$ as $3.82$ \unit{\angstrom} based on the bond length in the HPS-Urry model; neither this value nor the other constants in $V_0$ have any practical consequence for the performance of this metric since they are the same across all sequences.
As with $\tilde{R}_\mathrm{g}$ versus $R_\mathrm{g}$, we find that $\tilde{B}_2$ predicts phase behavior better than $B_2$ alone (Supporting Information, Figure S2) and thus use $\tilde{B}_2$ for all further calculations.

\paragraph{Expenditure density, $\rho^\mathrm{*}$.}
We introduce a new metric for phase-separation propensity, which we term the ``expenditure density'', or $\rho^\mathrm{*}$. 
Intuitively, $\rho^\mathrm{*}$ is the density to which a sequence can be reversibly compressed from infinite dilution given an allowance of work, $W^*$. 
Sequences with low phase-separation propensity will exhibit high pressures associated with intermolecular repulsion at lower densities and will quickly exhaust $W^*$.
For sequences with an increased phase-separation propensity, the same high pressures will not be experienced until higher densities and may even exhibit a negative pressure loop at intermediate densities, allowing for further compression.
Thus, higher $\rho^\mathrm{*}$ is associated with increased phase-separation propensity.

The expenditure density is calculated using the following equation 
\begin{equation}\label{eq:exp}
    \hat{w}^*=\frac{W^*}{m}=\int_0^{\rho^\mathrm{*}} d\rho\,\frac{P\left(\rho\right)}{\rho^2}
\end{equation}
where $\rho$ is the protein density, $P(\rho)$ is the pressure-density EoS (``Assessment of phase behavior'' in Section {\ref{subsec:sim_details}}), and $\hat{w}^*$, is the intensive work allowance.
Practically, $\hat{w}^*$ is a parameter that must be chosen to solve for $\rho^\mathrm{*}$.
We selected $\hat{w}^*=15$ \unit{atm\cdot\milli\liter\per\gram} to maximize the classification accuracy of $\rho^\mathrm{*}$, although the performance of $\rho^\mathrm{*}$ for classification is relatively robust as long as $\hat{w}^*$ is sufficiently large (Supporting Information, Figure S3).
To obtain $\rho^\mathrm{*}$ in practice, we numerically integrated the cubic spline for $P(\rho)$ and identified the density value that satisfies eq. \eqref{eq:exp}.
Standard errors for $\rho^\mathrm{*}$ were computed by constructing $10^4$ EoS using bootstrapped pressure values from the production run and recalculating $\rho^\mathrm{*}$.

\subsection{Machine Learning Details\label{subsec:ml_models}}
\paragraph{Sequence featurization.}
Each sequence was represented using an augmented scaled fingerprint~\cite{patel_data-driven_2023}, resulting in a 30-dimensional feature vector.
The features used are summarized in Table \ref{tab:sequence_features}.
The first twenty features are the composition of each residue type, $f_X$, where $X$ is the residue identity.
The last ten are the total number of residues, $N$, the fraction of positively and negatively charged residues, $f_+$ and $f_-$, the absolute value of the net charge per residue, $\overline{|q_i|}$, sequence charge decorator~\cite{firman_sequence_2018}, SCD, sequence hydropathy decorator~\cite{zheng_hydropathy_2020}, SHD, average hydrophobicity value per residue according to the Urry hydrophobicity scale, $\overline{\lambda_i}$, Shannon entropy of the sequence composition, $H$, average molecular weight per residue, $\overline{m_i}$, and the second virial coefficient based on a mean-field treatment, $B_2^\mathrm{MF}$.
Further details regarding the calculation of the features are provided in the Supporting Information, Section S4.
Throughout the remainder of the text, this 30-dimensional feature vector is referred to simply as a ``sequence feature vector.''

\begin{table}[t]
    \centering
    \caption{Summary of Sequence Features}
    \label{tab:sequence_features}
    \begin{tabularx}{\textwidth}{llcX}
        \hline
        \textbf{Category} & \textbf{Feature} & \textbf{Symbol} & \textbf{Description} \\
        \hline
        Composition & Amino acid fractions & $f_X$ & Fraction of each amino acid type \\
        & Sequence length & $N$ & Total number of residues \\
        & Molecular weight & $\overline{m_i}$ & Average molecular weight per residue \\
        \hline
        Charge & Positive charge fraction & $f_+$ & Fraction of basic residues (K, R) \\
        & Negative charge fraction & $f_-$ & Fraction of acidic residues (D, E) \\
        & Net charge per residue & $\overline{|q_i|}$ & Absolute value of net charge divided by length \\
        \hline
        Patterning & Charge decorator~\cite{firman_sequence_2018} & SCD & Charge distribution along sequence \\
        & Hydropathy decorator~\cite{zheng_hydropathy_2020} & SHD & Hydrophobic residue patterning \\
        \hline
        Derived & Average hydrophobicity & $\overline{\lambda_i}$ & Mean Urry hydrophobicity value \\
        & Sequence entropy & $H$ & Shannon entropy of composition \\
        & Mean-field $B_2$~\cite{an_active_2024} & $B_2^\mathrm{MF}$ & Theoretical prediction from composition \\
        \hline
    \end{tabularx}
\end{table}

\hfill

\paragraph{Dimensionality reduction.} We utilize dimensionality reduction by the uniform manifold approximation and projection (UMAP) algorithm\cite{mcinnes_umap_2020} to visualize the relationship between sequences. 
In essence, the UMAP algorithm identifies 
a low-dimensional embedding that effectively preserves the neighborhood structure of a higher-dimensional \textit{k}-neighbor graph.
In this case, the higher-dimensional graph is determined from 
the sequence feature vectors for all sequences ($\mathbf{X} \in \mathbb{R}^{2034 \times 30}$).
Prior to being used as input, the feature values all undergo standard scaling (i.e., removing the mean and scaling to unit variance) to be on the same unit scale.
Implementation is technically facilitated using the UMAP Python package with parameters set as `n\_neighbors=50', `min\_dist=0.5', and `n\_components=2'. 

\hfill

\paragraph{Threshold classifiers.} To assess whether a given metric can effectively predict phase behavior, we construct threshold classifiers and measure the area under the curve (AUC) of the receiver-operator characteristic. 
Threshold classification is performed by defining a threshold value for a metric beyond which phase separation is predicted to occur.
Given a threshold, $\theta$, sequences are categorized into one of four classification outcomes: true positive (TP: correctly classified phase-separating sequence), true negative (TN: correctly classified non-phase-separating sequence), false positive (FP: incorrectly classified non-phase-separating sequence), and false negative (FN: incorrectly classified phase-separating sequence).
These outcomes are used to compute  the true positive rate (TPR) and false positive rate (FPR) via
\begin{equation}
    \text{TPR}(\theta) = \frac{\text{TP}(\theta)}{\text{TP}(\theta) + \text{FN}(\theta)}
\end{equation}
and
\begin{equation}
    \text{FPR}(\theta) = \frac{\text{FP}(\theta)}{\text{FP}(\theta) + \text{TN}(\theta)}.
\end{equation}
The AUC is computed as
\begin{equation}
\text{AUC} = \int \text{TPR}(\theta)\left|\frac{d \, \text{FPR}(\theta)}{d \theta} \right| d\theta,
\end{equation}
which corresponds to numerically integrating the receiver operating characteristic curve.  
Since the choice of $\theta$ is somewhat arbitrary and application-specific, we use AUC as a more general metric for classification performance.

\hfill

\paragraph{Metric regressors.} Machine learning models based on neural networks (NNs) were used to predict phase-separation propensity metrics from sequence features. 
These predicted values were then supplied as input to threshold classifiers to predict whether a given IDP solution will phase separate, as determined by the presence of a negative pressure loop in its EoS (see ``Assessment of phase behavior'' in Section \ref{subsec:sim_details}); the fidelity of this model was then used to assess the utility of such metrics. 
All NNs had fully connected, feed-forward architectures with an input of the sequence feature vector and an output of the predicted value of a given metric.
All NNs were constructed using the scikit-learn~\cite{pedregosa_scikit-learn_2011}.
For NNs that predicted $\rho_\mathrm{c}$, $\rho^\mathrm{*}$, $\tilde{B}_2$, and $\tilde{R}_\mathrm{g}$, model training was treated as a regression task.
Non-phase-separating sequences (according to the EoS construction) were assigned a label of `0.0' for $\rho_\mathrm{c}$.
Standard scaling was applied to all features and labels during training of all machine learning models.

Models were trained on datasets ranging in size from 20 to 1,627 sequences to evaluate how performance varied with training set size. Training sequences were randomly selected without replacement, and the remaining sequences were used as the test set.
Model training employed the Adam solver~\cite{kingma_adam_2017}, with mean squared error used as the loss function. Each model was trained for up to $10^5$ epochs, using 20\% of the training data for validation. Training was stopped early if the validation loss failed to improve for ten consecutive epochs.
Hyperparameters were selected via grid search, using three-fold cross-validation on the training data. The combination yielding the highest average cross-validation score was chosen. Performance was evaluated using the area under the ROC curve (AUC) for classification tasks and the coefficient of determination ($R^2$) for regression. Details of the hyperparameter grid are provided in the Supporting Information, Section~S5.

Using the best identified combination of hyperparameters, ten models with different random weight initializations were trained. Final predictions on the test set were computed by averaging outputs across these ten models. This process was repeated 25 times for each metric and each training set size to estimate the mean AUC and its standard error. Statistical significance between AUC values was assessed using permutation testing with a significance threshold of $p < 0.01$.

\hfill

\paragraph{Direct classifiers.}
For comparison against models using the candidate phase-separation propensity metrics, NN models that directly predicted whether a sequence phase separates were also constructed. 
In this case, labels of 0 and 1 were assigned to non-phase-separating and phase-separating sequences, respectively.
In alignment with the discrete nature of the labels, the models used a logistic activation function for the output layer, which converts model outputs from a continuous space into a probability of having a label of 1.
The binary cross-entropy (BCE), which quantifies the difference between the predicted label probabilities and true binary labels, was used as the loss function instead of the mean squared error.
The BCE is given by
\begin{equation}
    \label{eq:bce}
    \text{BCE} = -\frac{1}{N} \sum_{i=1}^{N} \left[ y_i \log(\hat{y}_i) + (1 - y_i) \log(1 - \hat{y}_i) \right]
\end{equation}
where $y_i$ are the true labels 0 or 1 and $\hat{y}_i$ are the predicted probabilities of having the label 1 for all all sequences $i$ in a batch of size $N$.
All other procedures for model construction and training were identical to those just described in ``Metric regressors''.

\hfill

\paragraph{Prediction of condensed-phase diffusivity.}
We also investigated whether the candidate phase-separation propensity metrics can usefully inform prediction of properties outside of observed phase-separation. 
In particular, regression models were trained to predict the condensed-phase diffusivity of phase-separating sequences. 
These models consisted of fully connected neural networks that took as input the sequence feature vector, augmented with one of the individual metrics ($\rho_\mathrm{c}$, $\rho^\mathrm{*}$, $\tilde{B}_2$, or $\tilde{R}_\mathrm{g}$).
The regression models followed the same structure, training procedure, and hyperparameter selection process as in prior sections, with the following modifications. 
The labels for condensed-phase diffusivity were sourced from prior work.~\cite{an_active_2024,webb_thermodynamic_2023} Instead of using 25 random train–test splits, we employed five-fold cross-validation, with each fold serving as the test set once. All models used linear activation in the output layer and the mean squared error as the loss function. During hyperparameter selection, models were scored based on negative root mean squared error (RMSE).
After selecting hyperparameters for each fold, we trained ten models with randomly initialized weights using the chosen settings. Final predictions for each fold were obtained by averaging the outputs across these ten models, and RMSE was computed on the corresponding test set. The average RMSE and standard error were then calculated across all five folds. Statistical significance between RMSE values was evaluated using permutation testing with a significance threshold of $p < 0.05$.

\subsection{Analysis of Sequence Determinants\label{subsec:seq_determinants}}

\paragraph{Shapley additive explanations (SHAP) analysis.}
To identify key features driving predictions of phase-separation propensity metrics and phase behavior, we performed SHAP analysis~\cite{lundberg_unified_2017} on a series of neural network models.
For these, the model outputs were  $\rho^\mathrm{*}$, $\tilde{B}_2$, $\tilde{R}_\mathrm{g}$, and binary phase behavior, while the inputs consisted of the sequence feature vectors.
The surrogate models used for SHAP analysis followed the same architecture, training procedure, and hyperparameter selection as described for classification (see ``Metric regressors'' and ``Direct classifiers'' in Section \ref{subsec:ml_models}), with the following modifications. 
Instead of using 25 random train–test splits, we employed five-fold cross-validation, withholding one fold from training at a time. 
To ensure consistency in interpretation, values of $\tilde{R}_\mathrm{g}$ and $\tilde{B}_2$ were negated (multiplied by –1), so that higher values corresponded to greater phase-separation propensity.
After selecting hyperparameters for each fold, a single model was retrained using those settings and used to compute SHAP values for that fold. 
SHAP values were calculated using the Explainer object from the SHAP Python package~\cite{lundberg_unified_2017}. 
Since these models were used solely for interpretability, 
SHAP values were computed for all sequences, including those in the training set. 
Final SHAP values represent the average across the five folds.
SHAP values are in units of the model output and  remain in standard-scaled form; this facilitates comparison across metrics and with respect to sequence features.

\hfill

\paragraph{Feature analysis by metric-based rankings.}
To evaluate metric consistency, we carried out a pairwise ranking analysis.
In particular, we iteratively compare metric values for randomly selected pairs of sequences, with one being phase-separating and the other non-phase-separating.
A metric that is consistent in terms of observed phase behavior should report that the phase-separating sequence has a higher phase-separation-propensity.

In each iteration, one phase-separating (PS) and one non-phase-separating (NPS) sequence were sampled with replacement from the full dataset.
These sequences were then ranked according to the selected candidate metric.
For example, the sequence with the smaller $\tilde{R}_\mathrm{g}$ was deemed more likely to phase separate than the one with larger $\tilde{R}_\mathrm{g}$. 
When the metric placed the PS member ahead of the NPS member, the PS sequence was added to a ``correctly ordered PS'' (CO–PS) category, and the NPS sequence was added to a ``correctly ordered NPS'' (CO–NPS) category.
The opposite ranking generated contributions instead to  misordered PS (MO–PS) and misordered NPS (MO–NPS) categories. 
This process was repeated $10^5$ times.
Feature distributions formed from the populations within these four categories were then compared to identify sequence characteristics that correlate with successful or failed rankings by each metric. 
This analysis gauges the discriminatory power of the metrics and their overall consistency with observed phase-separation behavior, without relying on a trained classifier.

Differences between distributions were quantified by the Jensen–Shannon (JS) divergence, which is a symmetric measure of dissimilarity between probability distributions. 
The JS divergence between two discrete probability distributions with respect to a feature $x$,  $P(x)$ and $Q(x)$, is computed as
\begin{equation}
\text{JS}(P, Q) = \frac{1}{2} \sum_{x} P(x) \ln \frac{2P(x)}{P(x) + Q(x)} + \frac{1}{2} \sum_{x} Q(x) \ln \frac{2Q(x)}{P(x) + Q(x)}.
\label{eq:js-divergence}
\end{equation}
Here, the $P(x)$ and $Q(x)$ reflect the distributions generated from the CO-PS, CO-NPS, MO-PS, and MO-NPS.
Using eq. \eqref{eq:js-divergence}, JS$(P,Q)$ is bounded between 0 and $\ln(2) \approx 0.693$, with higher values indicating greater dissimilarity.
All probability distributions were represented by histograms with bin widths equal to one tenth of the standard deviation of the feature of interest. To reduce sensitivity to bin placement and discretization error, eq.~\eqref{eq:js-divergence} was evaluated ten times with histogram bin edges shifted incrementally by one tenth of a standard deviation. The resulting values were used to compute the mean and standard error of the JS divergence.

\section{Results \& Discussion\label{sec:results}}

\begin{figure}[t]
    \centering
    \includegraphics[width=0.9\textwidth]{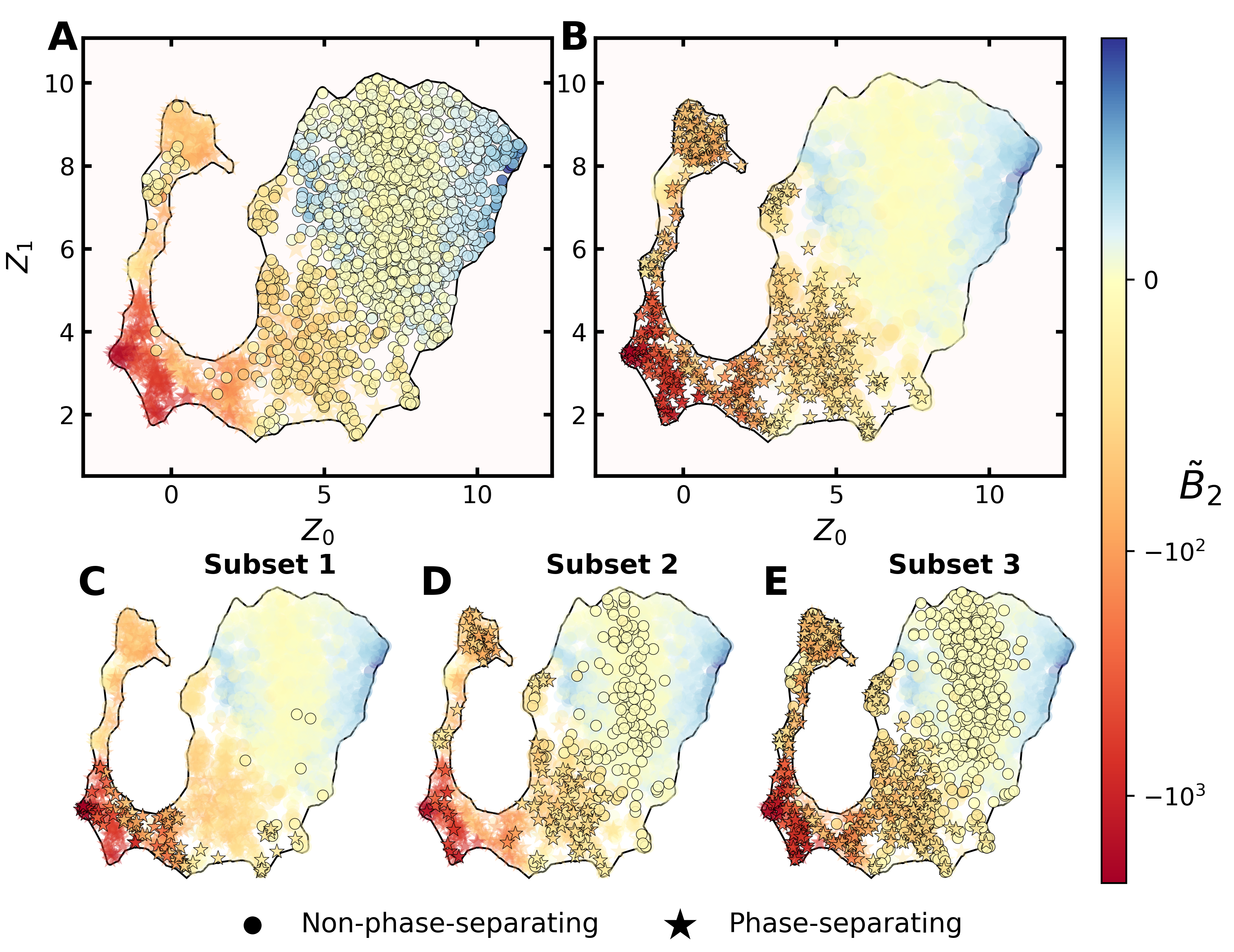}
    \caption{Visualization of phase behavior of studied sequences. A two-dimensional manifold organization of (\textbf{A}) non-phase-separating and (\textbf{B}) phase-separating sequences based on the Uniform Manifold Approximation and Projection (UMAP) unsupervised learning algorithm. The coordinates $Z_0$, $Z_1$ are obtained through a dimensionality reduction of the sequence feature vectors of all sequences (``Dimensionality Reduction'' in Section \ref{subsec:ml_models}). The distribution of sequences across the manifold conditioned for those featuring (\textbf{C}) no charged residues ($\mathcal{S}_1$), (\textbf{D}) charged residues but neutral chains ($\mathcal{S}_2$), and (\textbf{E}) net pairwise attraction and non-zero net charge ($\mathcal{S}_3$). The color bar (right) and marker legend (bottom) apply to all panels. Across all panels, sequences with the defined characteristics possess black outlines, while those with no outlines are shown for visual reference across panels.}
    \label{fig:main_fig}
\end{figure}

\subsection{Visualization of Sequence Characteristics\label{sec:characteristics}}

We begin by examining high-level relationships between the sequences, their features, and their association with phase separation. 
To do so, we visualize the studied sequence space by dimensionality reduction of the sequence feature vectors, projecting them onto a two-dimensional space using Uniform Manifold Approximation and Projection (UMAP)~\cite{mcinnes_umap_2020} (``Dimensionality reduction'' in Section \ref{subsec:ml_models}).
Figure~\ref{fig:main_fig} shows the resulting low-dimensional embedding, defined by coordinate axes $Z_0$ and $Z_1$. Each sequence is represented by a marker, with shape indicating whether the sequence undergoes phase separation and color denoting its dimensionless second virial coefficient, $\tilde{B}_2$.
By construction, sequences that are close in the low-dimensional embedding space also share similar sequence characteristics.

Figures \ref{fig:main_fig}A,B show that while non-phase-separating and phase-separating sequences generally occupy distinct regions of sequence space, there are also regions of considerable overlap between the groups.
Non-phase-separating sequences are particularly prevalent in the high-$Z_0$, high-$Z_1$ regions; these sequences also tend to possess near-neutral to positive $\tilde{B}_2$.
Meanwhile, most sequences with $Z_0\lesssim 2$ are predominantly phase-separating; this includes sequences with strongly negative $\tilde{B}_2$. 
However, there is a substantial band around intermediate $Z_0$, for which both phase-separating and non-phase-separating sequences are commonly encountered. 
The color annotation derived from $\tilde{B}_2$ in this area is similar across the two groups and extends from near-neutral to moderately negative $\tilde{B}_2$.
This overlap suggests how $\tilde{B}_2$ would be insufficient to classify the phase behavior of sequences and also how sequences of broadly similar sequence characteristics may possess disparate phase-separation behavior.~\cite{jin_predicting_2025} 

Figures~\ref{fig:main_fig}C–E illustrate how the low-dimensional embedding reflects sequence characteristics by highlighting three subsets of sequences. 
Each subset is defined by conditioning on specific sequence features or properties.
Figure~\ref{fig:main_fig}C shows the coordinates over a subset of sequences, labeled $\mathcal{S}_1$, which includes sequences with purely hydrophobic interactions ($f_+ = 0$, $f_- = 0$). 
These sequences are representative of those described by simple hydrophobic–polar models~\cite{statt_model_2020,rekhi_role_2023,rana_phase_2021}. 
They primarily cluster in the low-$Z_0$, low-$Z_1$ region of the embedding space and exhibit strongly negative $\tilde{B}_2$, with fewer points extending toward higher $\tilde{B}_2$ and non–phase-separating regions.
Figure~\ref{fig:main_fig}D displays sequences for a second subset, $\mathcal{S}_2$, comprising sequences with balanced but nonzero total charge ($f_+ = f_- > 0$). 
These mimic polyampholytic polymers~\cite{das_conformations_2013,lin_phase_2017,das_lattice_2018}, which were previously shown to phase separate while maintaining high chain mobility~\cite{an_active_2024}. 
These sequences generally have near-neutral to moderately negative $\tilde{B}_2$ and occupy the mixed region containing both phase-separating and non–phase-separating sequences.
Figure~\ref{fig:main_fig}E displays a final subset of sequences, $\mathcal{S}_3$, which includes sequences with net pairwise attraction and nonzero net charge ($\overline{|q_i|} > 0$, $\tilde{B}_2 < 0$). 
These sequences appear broadly distributed across the embedding space and were hypothesized to pose classification challenges due to competing effects of attraction and charge~\cite{mao_net_2010}.
Together, these subsets enable evaluation of metric performance across sequences with distinct characteristics.

\subsection{Analysis of Phase-separation Prediction Accuracy\label{sec:classification}}   

As a first assessment of the various phase-separation propensity metrics, we consider how well a given metric, when precisely known, can classify a sequence by its phase behavior at 300 K (see ``Threshold classifiers'' in Section \ref{subsec:ml_models}). 
Classifier performance was assessed using AUC, where a value of 1 indicates perfect classification and a value of 0.5 indicates naive classification (i.e., statistically indistinguishable from random guessing).

\begin{figure}[t]
    \centering
    \includegraphics[width=0.9\textwidth]{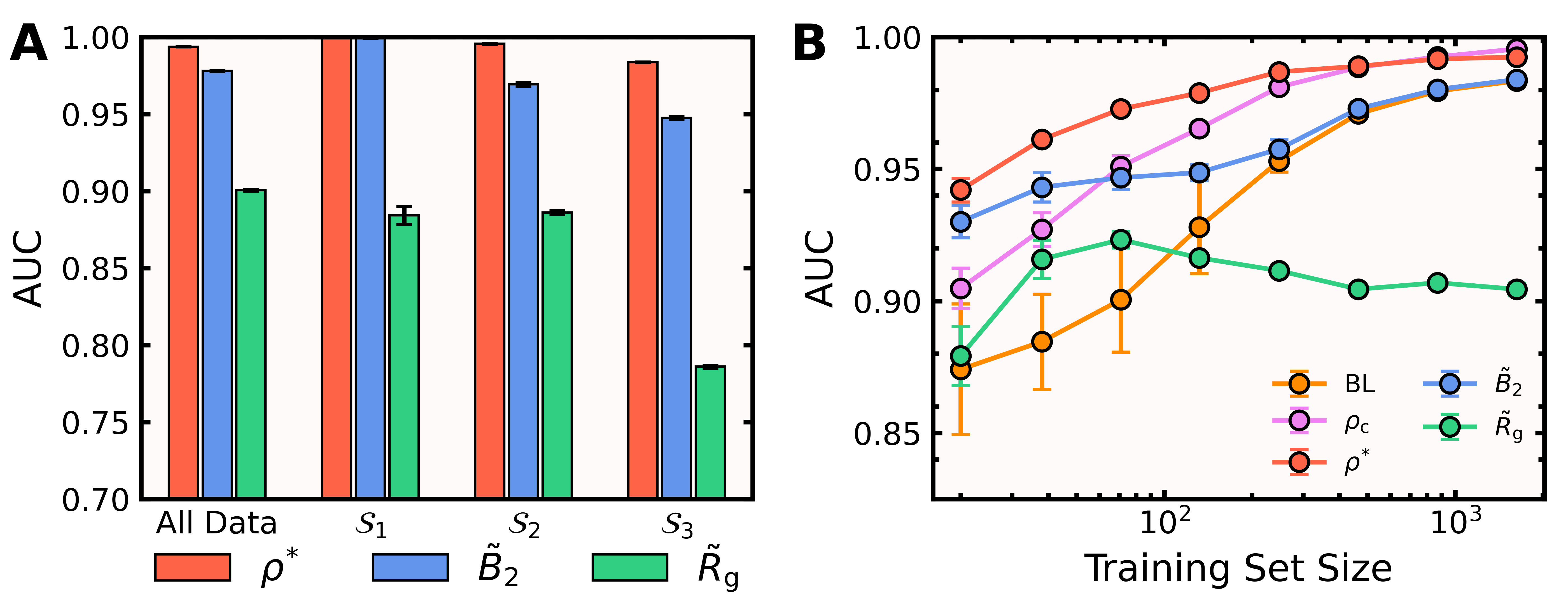}
    \caption{Comparison of classification accuracy for phase-separation propensity metrics. (\textbf{A}) Area under the receiver-operator characteristic curve (AUC) for threshold-based classification using each metric across different sequence populations. The leftmost grouping corresponds to evaluation over the entire dataset, while the remaining groupings highlight the subsets of sequences with conditional sequence features, as shown in Figures \ref{fig:main_fig}C-E. Error boxes represent the standard error calculated by bootstrapping metric values $10^3$ times. (\textbf{B}) Classification performance (based on AUC) of surrogate model classifiers as a function of training set size. In addition to the phase-separation propensity metrics--$\rho^\mathrm{*}$, $\tilde{B}_2$, and $\tilde{R}_\mathrm{g}$--results are also shown for classification based on a predicted condensed-phase density, $\rho_\text{c}$, and a simple binary label: BL = 1 (phase-separating) or BL = 0 (non-phase-separating). Error boxes represent the standard error across 25 random folds of data.}
    \label{fig:auc}
\end{figure}

Figure \ref{fig:auc}A compares metric performance on different populations of studied sequences. 
Evaluated on the whole dataset (left), $\rho^\mathrm{*}$ (AUC=0.994) outperforms $\tilde{B}_2$ (AUC=0.978), which in turn outperforms $\tilde{R}_\mathrm{g}$ (AUC=0.901).
While seemingly small differences, these AUC values indicate that $\tilde{B}_2$ and $\tilde{R}_\mathrm{g}$ are respectively 3.5 and 15.8 times more likely to misclassify a randomly chosen pair of phase-separating and non-phase-separating sequences than $\rho^\mathrm{*}$~\cite{hand_simple_2001}.
For sequences within $\mathcal{S}_1$ (left-middle), $\rho^\mathrm{*}$ and $\tilde{B}_2$ perfectly classify the subset of sequences (AUC=1), while the performance of $\tilde{R}_\mathrm{g}$ decreases relative to that on the whole dataset (AUC=0.884); this implies that predicting whether uncharged sequences will phase separate is more difficult on the basis of the $\tilde{R}_\mathrm{g}$ of a sequence compared to its $\rho^\mathrm{*}$ or $\tilde{B}_2$. 
For sequences within $\mathcal{S}_2$ (right-middle), $\rho^\mathrm{*}$ improves (AUC=0.996), while performance for $\tilde{B}_2$ (AUC=0.969) and $\tilde{R}_\mathrm{g}$ (AUC=0.886) worsens, relative to performance on the whole dataset.
For sequences within $\mathcal{S}_3$ (right), which was anticipated to represent a challenging set, performance decreases for all metrics (AUC=0.984, 0.947, and 0.786 for $\rho^\mathrm{*}$, $\tilde{B}_2$, $\tilde{R}_\mathrm{g}$).

The collection of results from Figure \ref{fig:auc}A demonstrates a clear, consistent hierarchy in performance where $\rho^\mathrm{*}$ ranks better or equivalent to  $\tilde{B}_2$, which outperforms $\tilde{R}_\mathrm{g}$.
We hypothesize that the observed performance hierarchy reflects the intrinsic capacity of each metric to capture multiscale interactions relevant to phase separation and to distill that information into a single scalar value.
By design, $\tilde{R}_\mathrm{g}$ captures only single-chain conformational behavior. While this may correlate with trends in many-chain systems, such alignment is neither guaranteed nor explicitly accounted for.
In contrast, $\tilde{B}_2$ characterizes pairwise interactions between chains and thus incorporates a higher level of interaction detail, which can enhance predictive performance.
Finally, $\rho^\mathrm{*}$ is derived from calculations that explicitly includes interactions between many chains, providing a more comprehensive representation of the collective behavior that governs phase separation.

\subsection{Analysis of Surrogate Model-based Classification and Data Efficiency\label{sec:efficiency}} 

Whereas the threshold classifiers analyzed in Fig.~\ref{fig:auc}A utilize the propensity metric values directly determined from simulation, in certain workflows, evaluating a metric for every candidate system may become impractical. 
Instead, models trained on existing data may be leveraged as efficient proxies for a given metric\cite{von_bulow_prediction_2025,lotthammer_direct_2024,von_bulow_machine_2025,patel_data-driven_2023}; predictions from these ``surrogate'' models can then go through subsequent threshold classification to predict phase behavior.
A simple strategy is to train the surrogate model to directly predict phase behavior based on binary observations.
In this case, predicted outputs are probabilities that can be subsequently mapped to binary labels using threshold classification. 
Nevertheless, the initial binary labels do not possess discriminative power within a given class (i.e., all phase-separating sequences have the same label of `1'). 

We hypothesized that using more physically informed and discriminating labels for the surrogate model, such as $\tilde{B}_2$ or $\tilde{R}_\mathrm{g}$, may improve the performance of classification, particularly in low-data regimes.
Therefore, we examine how well NN surrogate model-based classifiers, trained with varying amounts of data, can predict phase behavior from sequence features (see ``Metric regressors'' in Section \ref{subsec:ml_models}). 
For the surrogate model, we use the candidate propensity metrics as labels.
To complement comparisons among the primary candidate metrics, we also assess surrogate models trained on condensed-phase density, $\rho_\text{c}$, which provides continuous values for phase-separating sequences only, and the simple binary label, which offers no relative information on phase-separation propensity within each class.

Figure~\ref{fig:auc}B shows that surrogate models that are trained using different propensity metrics as labels indeed vary in their data efficiency when applied for classification of sequence phase separation.
In the limit of ``large'' training set sizes, the models separate into three performance tiers based on their training label.
The first tier consists of surrogate models using $\rho_\mathrm{c}$ and $\rho^\mathrm{*}$ as their training label, which achieve similarly high accuracy and both outperform models in the second tier, which use $\tilde{B}_2$ and the binary phase behavior. 
The performance in the last tier, featuring models with $\tilde{R}_\mathrm{g}$ as the training label, lags significantly behind the others.
In data-scarce scenarios (fewer than 200 samples), relative differences become more pronounced. 
Here, $\rho^\mathrm{*}$ emerges as the most predictive metric, followed by $\tilde{B}_2$ and $\rho_\mathrm{c}$, then $\tilde{R}_\mathrm{g}$ and the binary label.
In absolute terms, for any given metric, AUC increases as training set size increases; however, the performance of $\rho^\mathrm{*}$ is somewhat more robust yet consistently high relative to other metrics.
Strikingly, the performance hierarchy observed in Figure~\ref{fig:auc}A is preserved in Figure~\ref{fig:auc}B, even with as few as 20 training samples, which supports our motivating hypothesis.



We suggest that the observed performance differences among metrics are driven by two key factors. 
The first is whether the metric directly encodes information about phase behavior.
For example, a nonzero $\rho_\mathrm{c}$ confirms the existence of a condensed phase, whereas $\tilde{B}_2$ is only suggestive. 
The second is whether the metric provides a continuous measure across all sequences; $\rho^\mathrm{*}$ and $\tilde{B}_2$ can be computed for a sequence regardless of its phase behavior, while $\rho_\mathrm{c}$ is only defined for those that phase separate.
These results suggest that in high-data regimes, metrics containing complete phase behavior information are advantageous, even if they are not defined for all sequences. 
In contrast, in low-data regimes, continuous metrics are more beneficial, as they offer graded information on phase-separation tendency, even if not fully resolving the phase behavior.
Thus, metrics like $\rho^\mathrm{*}$ and $\tilde{B}_2$ may improve data efficiency in high-throughput learning of sequence–propensity relationships~\cite{an_active_2024,von_bulow_prediction_2025,lotthammer_direct_2024,tesei_conformational_2024}. Notably, $\rho^\mathrm{*}$ offers added utility despite being derived from the same simulations used to estimate phase behavior and compute $\rho_\mathrm{c}$.

\begin{figure}[t]
    \centering
    \includegraphics[width=0.45\textwidth]{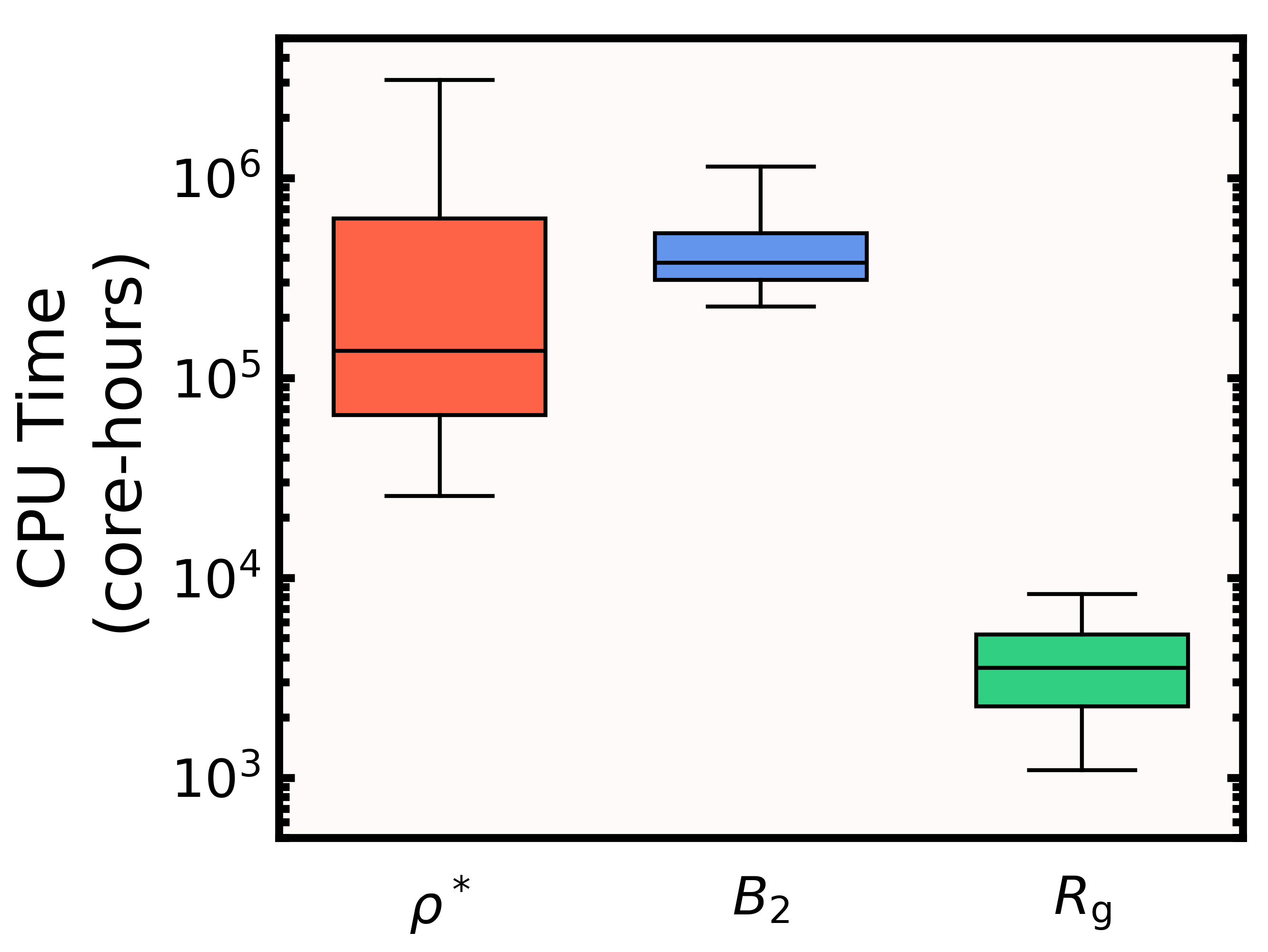}
    \caption{Comparison of computational cost across phase-separation propensity metrics. Distributions are shown for all 2,034 sequences using box-and-whisker plots. Boxes represent the interquartile range, with the horizontal line indicating the median. Whiskers denote the full range from minimum to maximum observed values. All reported numbers are based on simulations using 96-core nodes with 2.9 GHz Intel Cascade Lake processors. }
    \label{fig:cpu_time}
\end{figure}

\subsection{Additional Considerations for Metric Use\label{sec:computational_cost}}

The prior results establish the expenditure density as a high-quality metric to assess phase-separation propensity relative to perhaps less physically informed metrics, like $\tilde{B}_2$ and $\tilde{R}_\mathrm{g}$.
Another primary consideration for phase-separation propensity metrics would be their computational cost. 
Figure \ref{fig:cpu_time} provides insight into the relative computational investment of these candidate metrics. 
Among the three metrics, $\tilde{R}_\mathrm{g}$ is considerably less computationally intensive, with resource investment about one to two orders of magnitude less than $\rho^\mathrm{*}$ and $\tilde{B}_2$; however, it is also the least accurate. 
Meanwhile, the computational cost of obtaining $\rho^\mathrm{*}$ or $\tilde{B}_2$ is comparable.
One difference is that the cost of computing $\rho^\mathrm{*}$ is more varied than $\tilde{B}_2$, although the former possesses a lower median and minimum cost.
This is because calculating $\rho^\mathrm{*}$ only requires simulations at densities up to and slightly beyond the value of $\rho^\mathrm{*}$, which can be estimated from simulations at lower densities, thus eliminating the need to explicitly examine the entire density range.
By contrast, the cost of calculating $\tilde{B}_2$ and $\tilde{R}_\mathrm{g}$ is more consistent and mainly scales with sequence length.
We note that some sequences occasionally required substantial investment to obtain $\rho^\mathrm{*}$, even exceeding $10^6$ CPU-hours.
Such sequences are typified by many charged residues and a larger $\rho^\mathrm{*}$, thus requiring simulations at higher densities and leading to more force evaluations within the large electrostatic cutoff distance. 
It is possible that such factors could be alleviated by using a different work allowance or by otherwise enhancing algorithmic efficiency.

Ultimately, selecting an appropriate phase-separation propensity metric will depend on numerous factors. 
Because computational investment is similar, it may be worthwhile to consider that accessing $\rho^\mathrm{*}$ is technically less involved than computing $\tilde{B}_2$.
Whereas the former depends only on sampling pressures from the canonical ensemble, the latter usually requires application of enhanced sampling or free-energy methods.\cite{comer_adaptive_2015,R:2018_SSAGES_JCP}
Nevertheless, the second virial coefficient is a widely accepted and well-defined quantity, which may motivate its use despite its moderate predictive performance. Similarly, $\tilde{R}_\mathrm{g}$ may be the metric of choice when computational resources are limited, given its low cost. 
As currently defined, eq. {\eqref{eq:exp}} for $\rho^\mathrm{*}$ is straightforwardly applied to models featuring implicit solvent, but extension to explicit-solvent simulations requires a more nuanced interpretation.
Moreover, $\rho^\mathrm{*}$ would seem non-trivial to readily connect to an experimental observable, apart from the assessment of phase separation. However, when feasible, $\rho^\mathrm{*}$ offers a favorable balance between accuracy and computational efficiency, making it a strong candidate for applications that aim to screen or model phase-separation propensity with limited data.

\begin{figure}[t]
    \centering
    \includegraphics[width=0.45\textwidth]{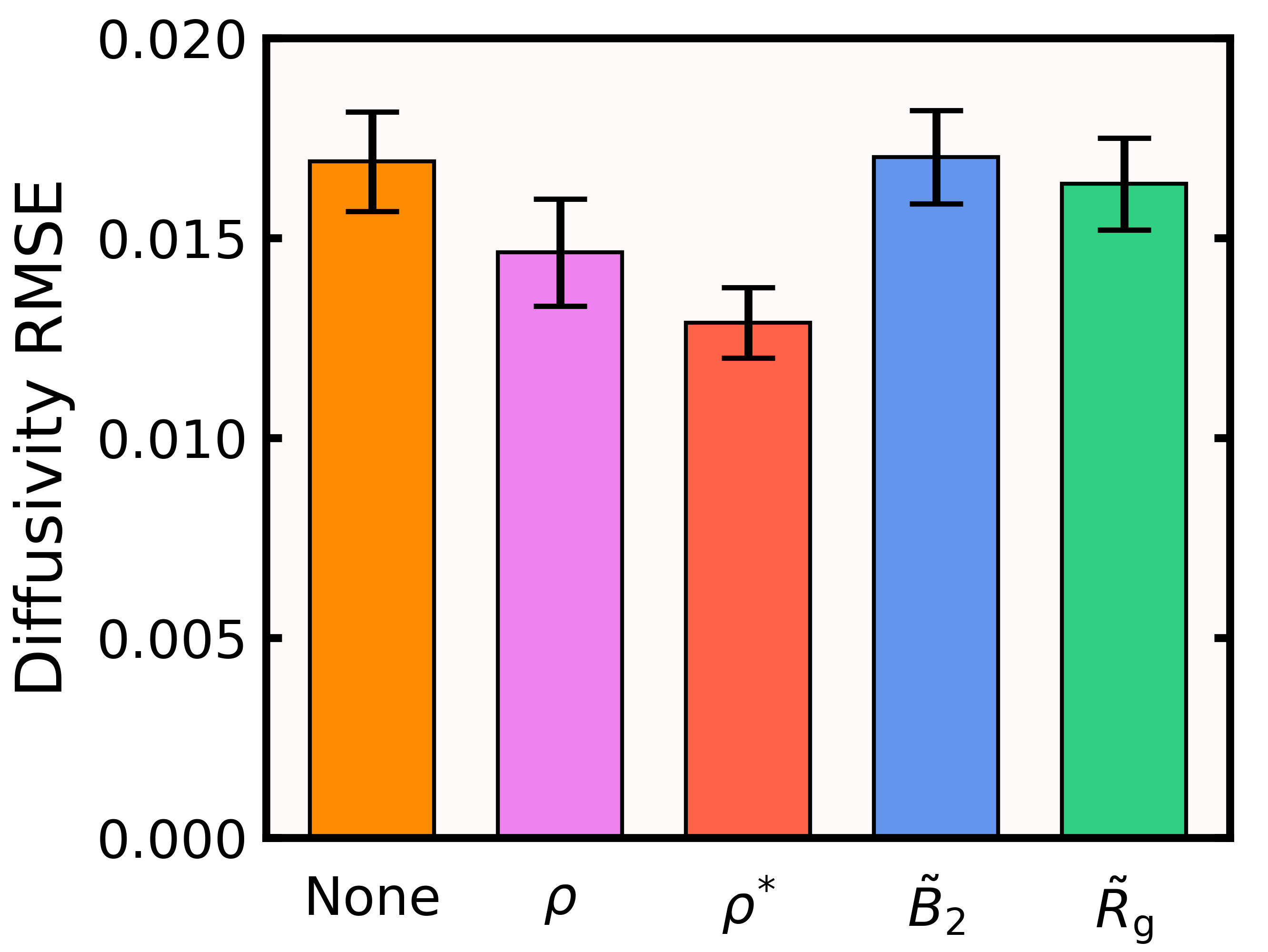}
    \caption{Performance on prediction of sequence condensed-phase diffusivity utilizing metrics as features. Bar chart showing the root mean squared error of NN-based predictions of sequence condensed-phase diffusivity for all phase-separating sequences (``Prediction of condensed-phase diffusivity'' in Section \ref{subsec:ml_models}), where the NN input is the sequence feature vector alone and augmented individually with $\rho_\mathrm{c}$, $\tilde{B}_2$, $\tilde{R}_\mathrm{g}$, and $\rho^\mathrm{*}$. Error boxes represent the standard error across 5-fold cross-validation.}
    \label{fig:diffusivity_errors}
\end{figure}

\subsection{Application of Metrics to Off-target Property Prediction\label{sec:diffusivity_prediction}}

Because the phase-separation propensity metrics capture information about underlying physical interactions, we also examined their utility for predicting physical properties not directly related to phase behavior. 
To this end, we trained neural network regression models to predict the self-diffusivity of sequences in the condensed phase from sequence feature vectors individually augmented with one of the following metrics: $\rho_\mathrm{c}$, $\tilde{B}_2$, $\tilde{R}_\mathrm{g}$, or $\rho^\mathrm{*}$.
This task necessarily only pertains to phase-separating sequences, such that augmenting the sequence feature vector with its binary phase behavior is equivalent to having no additional feature (see also ``Prediction of condensed-phase diffusivity'' in Section \ref{subsec:ml_models}). 

Figure \ref{fig:diffusivity_errors} shows that only $\rho^\mathrm{*}$ provides a statistically significant reduction in the error of diffusivity predictions when compared to having no additional feature.
Although not statistically significant, the next largest reduction in prediction error is observed with $\rho_\text{c}$.
Compared to the metrics based on single- or two-chain interactions, this seems reasonable given that this metric directly reflects conditions of the condensed phase, and one might generally expect chain diffusivity to be inversely related to density based on free-volume considerations.
Given this, it is intriguing that $\rho^\mathrm{*}$ appears to outperform $\rho_\text{c}$, though confirming such would require additional sampling and data. 
Nonetheless, these results indicate that $\rho^\mathrm{*}$ could be used to more broadly inform the construction of sequence–property relationships for IDPs as a general descriptor of many-chain interactions.

\subsection{Sequence Determinants of Candidate Metrics\label{sec:feature_analysis}}
\begin{figure}[tb]
    \centering
    \includegraphics[width=0.9\textwidth]{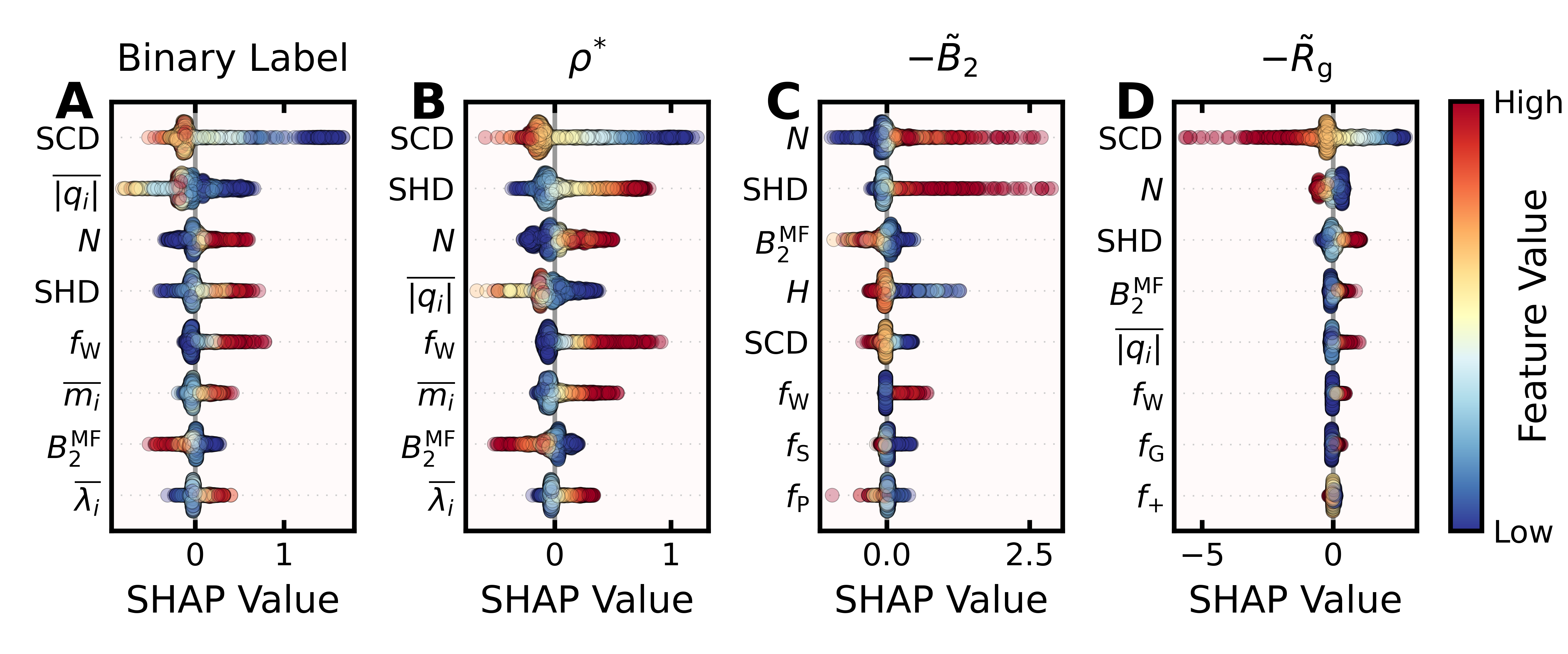}
    \caption{Feature importance analysis for phase behavior and metrics of propensity. Beeswarm plots of SHAP values (``SHAP analysis'' in Section \ref{subsec:seq_determinants}) for the top-8 most influential features for predictions made by NN surrogate models on (\textbf{A}) binary phase behavior, (\textbf{B}) $\rho^\mathrm{*}$, (\textbf{C}) $-\tilde{B}_2$, and (\textbf{D}) $-\tilde{R}_\mathrm{g}$. Each marker corresponds to a prediction for a specific sequence. Features are ordered from the top of each chart by average absolute SHAP value across all sequences. Only the top eight features are shown for each model. A larger vertical distribution of points represents greater density of points around that value. Points are colored by relative feature value. Model predictions are transformed by standard scaling and signs are adjusted so that positive SHAP values correspond with increased phase-separation propensity to aid in comparison between panels.}
    \label{fig:shap}
\end{figure}

To better understand each metric and its relationship to phase separation, we performed a feature-importance analysis using SHAP (``SHAP analysis'' in Section \ref{subsec:seq_determinants}). 
SHAP analysis assigns values for each feature that reflect both the direction and magnitude of its contribution to a given prediction. 
For instance, if the SHAP value for SCD of a given sequence is large and positive, and the SCD of that sequence is low, that implies that the low charge patterning enhanced $\rho^\mathrm{*}$ for the prediction of that sequence.
This analysis can be examined at a population-level to identify major trends.
Furthermore, ranking features by average absolute SHAP values provides a means to assess the overall importance of features to model predictions. 
For high-quality surrogate models, this importance may reflect the relevance of the underlying physical quantity.
Figure \ref{fig:shap} summarizes how individual components of the sequence feature vector influence predictions of phase behavior directly compared to their influence on predictions of phase-separation propensity metrics.
To make the interpretation of SHAP values consistent across all candidate metrics, feature influence was assessed for $-\tilde{R}_\text{g}$ and $-\tilde{B}_2$.
In this way, positive SHAP values correspond to increased likelihood of phase separation, for all metrics.

As a baseline, we find that the sequence determinants of phase separation broadly align with expectations from the existing literature (Figure \ref{fig:shap}A).
To begin, SCD is ranked as the most influential feature across this dataset in predicting whether a sequence will phase separate, with the identifiable trend that lower SCD (cool colors) tends to promote phase separation (positive SHAP) while higher SCD (warm colors) suppresses this behavior (negative SHAP).
This is consistent with observations that sequences with a more blocky charge distribution are more likely to phase separate than those with more distributed charges~\cite{von_bulow_prediction_2025,lin_phase_2017,dignon_temperature-controlled_2019,tesei_conformational_2024,mccarty_complete_2019}.
The average net charge per residue is the next most influential feature, with highly charged sequences (warm colors) being less likely to phase separate (negative SHAP), and following this feature is overall sequence length, for which longer sequences (warm colors) are more likely to phase separate.
These trends have also been previously noted~\cite{dignon_sequence_2018,mao_net_2010,tesei_conformational_2024,von_bulow_prediction_2025}.
The features that follow are more directly related to hydrophobicity where enhanced hydrophobicity (SHD, $f_\mathrm{W}$, $\overline{\lambda_i}$) promotes phase separation; W (tryptophan) is the most hydrophobic residue in the Urry scale.
According to feature rankings, SHD, which captures facets of how hydrophobicity is distributed on a chain, is more influential than the fraction of tryptophan ($f_\text{W}$) or average hydrophobicity ($\overline{\lambda_i}$). 
This trend aligns with previous observations in the literature.\cite{tesei_conformational_2024,von_bulow_prediction_2025,zheng_hydropathy_2020}

It is important to note that that SHAP analysis does not account for interactions or correlations between features, such that apparent feature importance may arise from indirect effects or covariates; this can make it difficult to unambiguously identify the true drivers of a prediction.
Despite this, the consistency with prior studies supports the utility of this analysis. 
In addition, it is intriguing that the observed trends emerge  even though our dataset includes significant populations of both biological and abiological sequences. This suggests that principles derived from synthetic or non-genomic sequences~\cite{pesce_design_2024} may offer insights into biologically relevant behavior and vice versa.
We now consider to what extent the deduced important features for phase-separation propensity metrics align with those identified as influential for predicting phase behavior. 
Our general hypothesis is that there would be consistency, particularly for more accurate metrics of phase-separation propensity, while any differences may highlight possible deficiencies in proposed metrics.

Comparing Figures \ref{fig:shap}A,B reveals remarkable similarity--in terms of relative importance, magnitude, and direction--for SHAP analysis applied to surrogate models of phase separation and those for $\rho^\mathrm{*}$.
Both indicate the same top-eight features and in the same order of ranked importance, except for a swap between SHD (ranked second) and $\overline{|q_i|}$ (ranked fourth), though both remain in the top four and behave similarly.
Considering $\tilde{B}_2$ (Figure \ref{fig:shap}C), the strongest determinants are SHD (ranked first) and length (ranked second), with relatively low importance assigned to charge-related features such as SCD (ranked fifth) and $\overline{|q_i|}$ (unranked) when compared to the binary label of phase separation.
This suggests that $\tilde{B}_2$ is influenced more strongly by hydrophobicity-related features and perhaps fails to adequately capture the effects of charge patterning, which appear more influential for predicting phase behavior; a notable omission from its top-eight features is average net charge per residue.
Finally, for $\tilde{R}_\mathrm{g}$ (Figure~\ref{fig:shap}D), some top features overlap with those important for predicting phase behavior directly. 
However, the overall magnitudes and patterns of influence differ substantially. Among all features, only SCD shows a noticeably different level of influence, suggesting that most features have limited or subtle effects on $\tilde{R}_\mathrm{g}$ predictions.
Moreover, certain features exhibit opposing trends compared to phase behavior. 
For example, higher values of $\overline{|q_i|}$ tend to correspond to more negative $\tilde{R}_\text{g}$ (i.e., more compact chains), while lower $\overline{|q_i|}$ values are associated with increased phase-separation propensity in Figure~\ref{fig:shap}A.
Taken together, these results suggest that $\rho^\mathrm{*}$ performs well as a classifier precisely because its sequence determinants closely align with those of phase separation and captures core physical principles. 

\subsection{Assessment of Metric Consistency with Observed Phase Behavior Outcomes\label{subsec:failure}}

We further elucidate metric performance by examining sequence features conditioned on the ability of a given metric to rank sequences in alignment with their phase behavior (see ``Feature analysis by metric-based rankings'' in Section \ref{subsec:seq_determinants}). 
Specifically, we generate bootstrapped populations of correctly ordered (CO) and misordered (MO) sequences and further categorize them by phase-separating (PS) and non-phase-separating (NPS) to generate four groups: CO-PS, CO-NPS, MO-PS, and MO-NPS.
The general expectation is that these four groups will resemble populations of TP, TN, FP, and FN categories generated from a classifier.
This analysis can identify whether certain features are enriched or depleted in these subpopulations relative to the overall dataset, with the rationale that features consistently present in correctly ordered or misordered sequences may reveal systematic limitations in the candidate metrics.

\begin{figure}[htb]
    \centering
    \includegraphics[]{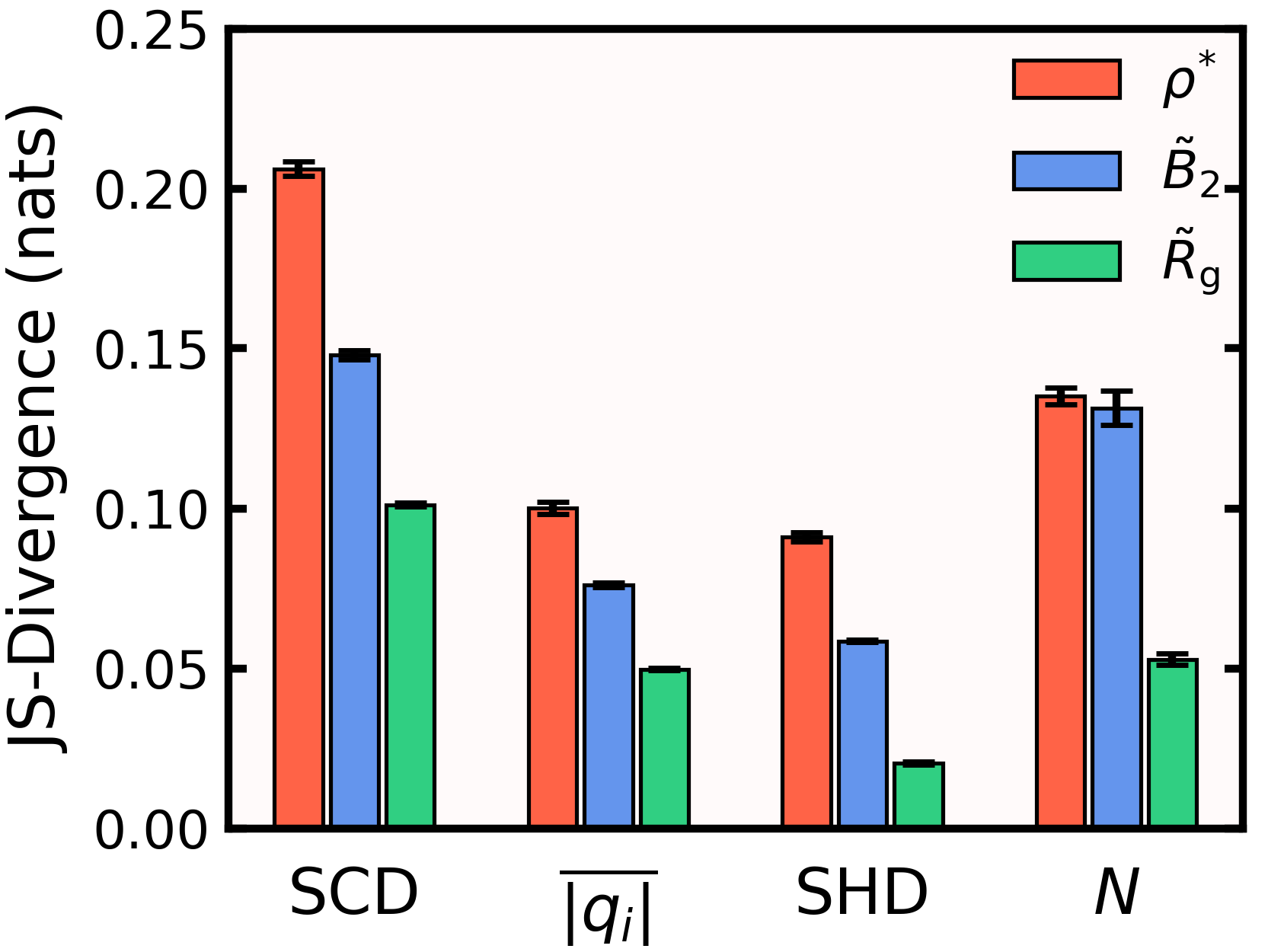}
    \caption{Feature differences between sequences conditioned on ranking outcome. Bar chart of the JS-divergence between the feature distributions of MO sequences (MO-PS and NO-NPS) and random sequences (CO-PS, CO-NPS, MO-PS, and MO-NPS) for the four features with the highest average magnitude of SHAP values for phase behavior when ranked by $\rho^\mathrm{*}$, $\tilde{B}_2$, and $\tilde{R}_\mathrm{g}$. Larger divergence values indicate greater differences between the distributions. JS-divergence was calculated using the natural logarithm, which bounds it between 0 and 0.693. Error boxes represent the standard error across 10 instances shifting the histogram bin edges used for construction of the feature distributions by one tenth of a standard deviation of the feature to reduce sensitivity to bin placement (see ``Feature analysis by metric-based rankings'' in Section \ref{subsec:seq_determinants}).}
    \label{fig:js-div}
\end{figure}

Figure \ref{fig:js-div} examines the prevalence of certain features (SCD, $\overline{|q_i|}$, SHD, and $N$) in populations of MO sequences (both MO-PS and MO-NPS) by a given phase-separation propensity metric.
This analysis uses the Jensen–Shannon (JS) divergence (see eq.~\eqref{eq:js-divergence}) to quantify how feature distributions in MO sequences differ from those in a randomly sampled background population;
higher JS divergence indicates greater dissimilarity between the two distributions. 
In this context, a larger JS divergence for a given feature suggests that the feature is more strongly associated with MO sequences than with sequences randomly drawn from the full dataset.

Across the analyzed features, which are the top-four most influential for phase behavior in Figure~\ref{fig:shap}A, the JS divergence is consistently highest for $\rho^\mathrm{*}$, followed by $\tilde{B}_2$ and then $\tilde{R}_\mathrm{g}$.
This striking progression indicates that MO sequences ranked by $\rho^\mathrm{*}$ are more distinct in their feature profiles compared to the overall sequence population, whereas MO sequences ranked by $\tilde{R}_\mathrm{g}$ more closely resemble the general distribution.
Comparing JS divergence magnitudes across features reveals that differences in SCD and $N$ are more prominent than $\overline{|q_i|}$ or SHD, suggesting that MO sequences more commonly arise when sequences are more unique in terms of these characteristics.
We interpret this to mean that reliable metrics like $\rho^\mathrm{*}$ tend to improperly rank sequences that are atypical or extreme relative to the dataset, while misorderings by less reliable metrics such as $\tilde{R}_\mathrm{g}$ are less specific in origin.
We note that $\rho^\mathrm{*}$ does not intrinsically possess information on sequence length, while its failure modes are reflected by sequence features that include this information (explicitly or implicitly).
Therefore, incorporating explicit length information may improve performance, as might targeted data acquisition in regions of feature space characterized by extreme sequence properties.

\begin{figure}[t]
    \centering
    \includegraphics[width=0.9\textwidth]{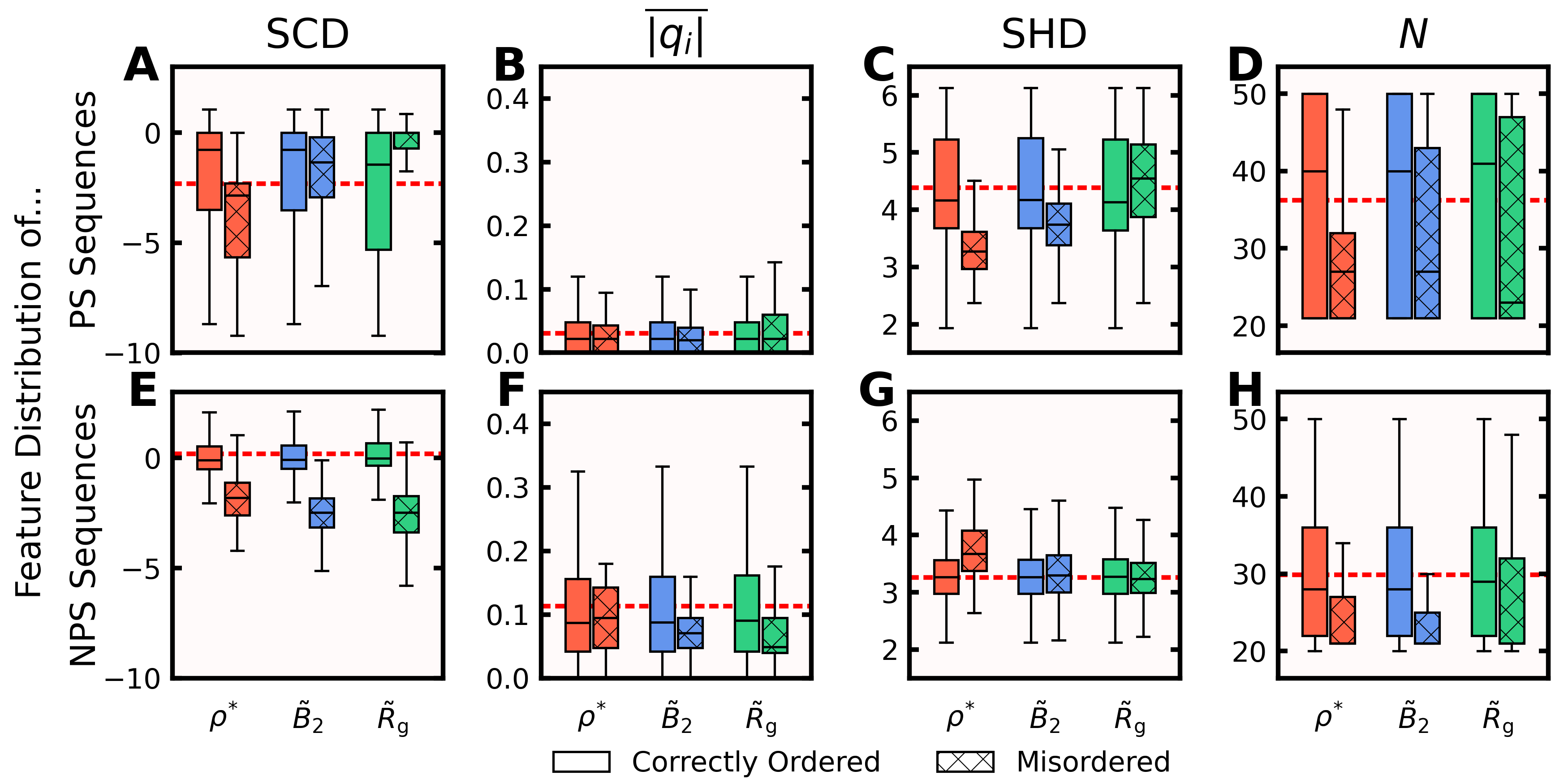}
    \caption{Feature distributions of sequences conditioned on the ability of a propensity metric to rank sequences in alignment with phase behavior (see ``Feature analysis by metric-based rankings'' in Section \ref{subsec:seq_determinants}). 
    For phase-separating (PS) sequences, boxplots for the distributions of (\textbf{A}) SCD, (\textbf{B}) $\overline{|q_i|}$, (\textbf{C}) SHD, and (\textbf{D}) $N$. 
    Within each panel, distributions are further conditioned by whether sequences are correctly ordered or misordered when ranked by a given metric.
    The box represents the inner quartile range and median values while the whiskers show the nearer to the quartiles of 1.5 times the inner quartile range or the maximum/minimum value.
    Panels (\textbf{E}-\textbf{H}) follow the same as (\textbf{A}-\textbf{D}), except applied for non-phase-separating (NPS) sequences.
    The dashed, horizontal red line indicates the mean of the given feature for the entire population of sequences with the stated phase behavior.
    In panel \textbf{A}, the apparent absence of a median line for misorderings by $\tilde{R}_\text{g}$ is due to equivalence in values for the second and third quartiles due to the number of neutral sequences, yielding SCD = 0. 
    In panel \textbf{B}, the apparent absence of median lines for misorderings is due equivalence in values for the first and second quartiles due to the number of sequences with $N=21$. 
    }
    \label{fig:features}
\end{figure}

Figure \ref{fig:features} compares the feature distributions of CO and MO sequences for both PS and NPS sequences to pinpoint feature determinants that drive correct and incorrect rankings vis-\`{a}-vis phase behavior.
This allows us to explicitly examine the directionality of changes in sequence features between CO and MO sequences while simultaneously conditioning our analysis on the type of misordering.
Unsurprisingly, the features of the populations of CO sequences correlate with their known impact on phase-separation propensity.
CO-PS sequences have lower SCD (blocky charge), lower $\overline{|q_i|}$, higher SHD (highly hydrophobic), and higher $N$ (Figures \ref{fig:features}A-D). 
CO-NPS sequences have the opposite characteristics (Figures \ref{fig:features}E-H).
Notably, the feature distributions of CO sequences are broadly similar across the different phase-separation propensity metrics (plain boxes). 
This suggests that sequences correctly ranked vis-\`{a}-vis phase behavior, or not, tend to share common sequence features regardless of the metric used, or possibly that any reasonable metric may be sufficient to correctly rank sequences with these characteristics.
The feature distributions of MO sequences (hatched boxes), however, are unique to each metric and provide insight into their shortcomings.

We find that phase-separation propensity metrics fail in distinct ways for MO-PS sequences (misorderings/hashed boxes across Figures \ref{fig:features}A-D).
For $\tilde{R}_\mathrm{g}$, MO-PS sequences typically involve sequences with high SHD, high SCD, and low $\overline{|q_i|}$. 
This suggests that strong hydrophobic interactions can drive phase separation in the absence of net charge, while a distributed charge patterning prevents chain collapse resulting in a larger radius of gyration.
For $\tilde{B}_2$, more frequent misorderings are associated with sequences with moderate SHD and SCD, and low $\overline{|q_i|}$. 
These sequences combine hydrophobicity with distributed charges, producing weakly attractive, nearly neutral chains that still undergo phase separation.
In contrast, $\rho^\mathrm{*}$ tends to underestimate phase separation for sequences with low SHD, low SCD, and low $\overline{|q_i|}$. 
This pattern suggests highly blocky, charge-neutral sequences with limited hydrophobicity, yet sufficient multichain interactions to support phase separation.

By contrast, all three metrics exhibit a common failure mode for MO-NPS sequences (misorderings/hashed boxes across Figures \ref{fig:features}E-H).
In particular, the likelihood of ranking a sequence as phase separating is overestimated for shorter sequences with more negative SCD and lesser $\overline{|q_i|}$ relative to the population expectations.
In these cases, blocky charge distributions lead to compact single-chain conformations and favorable two- or even many-chain interactions that can explain their misordering.
However, bulk phase separation is precluded by a net charge accumulation that cannot be overcome by weak hydrophobic interactions.
We hypothesize that sequences with these characteristics form finite-size aggregates, but aggregate formation is driven by charge accumulation, rather than the formation of micelles with a hydrophobic core and hydrophilic exterior as observed in some model IDP systems~\cite{statt_model_2020,rana_phase_2021,rekhi_role_2023}.
These systematic under- and over-predictions, combined with the observation that MO sequences tend to be shorter than their CO counterparts, highlight possible areas for improving future classification approaches.

\section{Conclusions\label{sec:conclusion}} 

We evaluated the efficacy of simple metrics for quantifying the propensity of intrinsically disordered proteins (IDPs) to undergo phase separation. 
Two of the metrics, the normalized radius of gyration ($\tilde{R}_\mathrm{g}$) and the normalized second virial coefficient ($\tilde{B}_2$), were based on conventional approaches and reflect simple expectations from respectively single-chain and two-chain physics. 
A third, newly proposed metric, the expenditure density ($\rho^\mathrm{*}$), instead links system interactions to phase behavior by integrating an approximate equation of state as part of a computation of boundary ($p-V$) work.
Evaluation of these metrics was facilitated using coarse-grained molecular dynamics simulations and machine learning for 2,034 chemically suggestive models of IDPs, including sequences with biological and abiological origins.

Our results demonstrated a clear performance hierarchy in terms of predicting sequence phase behavior ($\rho^\mathrm{*} > \tilde{B}_2 > \tilde{R}_\mathrm{g}$).
The simple rationale for this progression aligns with the complexity of interactions distilled by each.
The expenditure density ($\rho^\mathrm{*}$), derived from simulations across a range of conditions, inherently incorporates many-chain interactions. 
In contrast, $\tilde{B}_2$ and $\tilde{R}_\mathrm{g}$ are limited to capturing two-body and single-chain interactions, which are less directly tied to macroscopic phase behavior in certain scenarios.
As such, $\rho^\mathrm{*}$ better captures key physical factors underlying phase separation, which was further supported by feature-importance analysis. 
This revealed that the most influential features (e.g., charge and hydrophobic patterning, sequence length, net charge) for predicting $\rho^\mathrm{*}$, as well as the direction and magnitude of their effect, closely mirrored those observed for predicting phase behavior directly.
In contrast, the same analysis revealed likely failure modes for other metrics. For example, $\tilde{B}_2$ tended to overemphasize hydrophobic interactions while underrepresenting electrostatic effects.

Our results also highlighted several prospective advantages of $\rho^\mathrm{*}$ beyond its strong classification accuracy. 
We found that $\rho^\mathrm{*}$ was more data-efficient based on it attaining higher accuracy with smaller training sets relative to other metrics. 
This advantage stems not only from its alignment with the underlying physics of phase behavior but also from its applicability to all sequences; this is unlike condensed-phase density, which provides graded information only for phase-separating sequences.
In terms of computational cost, $\rho^\mathrm{*}$ was comparable to the second virial coefficient but arguably technically simpler to compute, as it does not require enhanced sampling or free-energy methods. These properties make $\rho^\mathrm{*}$ especially well-suited for data-driven screening and machine learning applications.

Overall, $\rho^\mathrm{*}$ offers a valuable addition to the computational toolkit for studying biomolecular phase separation. 
Despite the clear performance hierarchy among the evaluated metrics, none achieved perfect classification accuracy. 
A deeper understanding of their shortcomings may help guide the development of better training datasets and improved metrics. 
All metrics had the ability to properly order sequences with respect to their phase behavior when sequence features correlated with their known impact on phase-separation propensity; however, misorderings were unique to each metric.
For misordered species broadly, errors by $\rho^\mathrm{*}$ were concentrated among sequences with more extreme features, while those for $\tilde{B}_2$ and $\tilde{R}_\mathrm{g}$ were more diffusely distributed. 
All metrics shared a common tendency to misorder non-phase-separating sequences characterized by more negative SCD, moderate net charge, and short sequence lengths. 

These findings also complement our recent work,\cite{jin_predicting_2025} which showed how statistical features extracted from two-chain contact-map simulations could significantly improve predictions of phase behavior compared to both $R_\text{g}$ and $B_2$ alone. 
Compared to $B_2$, the contact-map approach is similar in computational expense but provides richer structural detail about pairwise interchain interactions.
In contrast, the expenditure density ($\rho^\mathrm{*}$) lacks information about structural correlations but does capture collective phenomena beyond pairwise physics; it is also technically straightforward. 
Thus, contact-map statistics and the expenditure density represent complementary approaches, balancing detailed pairwise structural information against computationally accessible multichain physics.

We envision multiple directions to extend from the current work.
Improving classifications by $\rho^\mathrm{*}$ may be tractable, as its errors appear more systematic and confined to specific regions of feature space.
In this, further understanding metric-specific limitations will be useful for curating robust training sets and generating effective models.
While our analysis focused on single-component systems under fixed conditions, the underlying framework provides a foundation for extending these approaches to more complex, biologically relevant scenarios, including multicomponent condensates and variable solution environments~\cite{saar_theoretical_2023}.
It may be useful to investigate whether contact maps obtained from simulations at densities near or within condensed phases (as required for calculating $\rho^\mathrm{*}$) could provide additional structural insights, facilitating an integrated approach that combines computational efficiency with detailed interaction information.
Finally, $\rho^\mathrm{*}$ may be used to study phase behavior in systems beyond biomolecular condensates, and extending its use to explicit-solvent systems is another area currently under investigation.

\begin{acknowledgement}
W.W.O. and M.A.W. acknowledge support from the National Science Foundation under Grant No. 2237470. 
Simulations and analyses were performed using resources from Princeton Research Computing at Princeton University, a consortium led by the Princeton Institute for Computational Science and Engineering (PICSciE) and Office of Information Technology's Research Computing. The authors also acknowledge Jessica Jin and Dr. Zachary Lipel for helpful discussions. 
\end{acknowledgement}

\section*{Data availability}
The data associated with this study are publicly accessible at \url{https://github.com/webbtheosim/ps-propensity-metrics}. 
Additional files to facilitate reproduction of simulations are available at 
\url{https://github.com/webbtheosim/md-simulation-files/tree/main/2025-idp-psp}.

\section*{Supporting Information}
Example Equations-of-State; Effect of Normalization on Metric Classification Accuracy; Expenditure Density Sensitivity Analysis; Sequence Featurization; Hyperparameter Grid

\providecommand{\latin}[1]{#1}
\makeatletter
\providecommand{\doi}
  {\begingroup\let\do\@makeother\dospecials
  \catcode`\{=1 \catcode`\}=2 \doi@aux}
\providecommand{\doi@aux}[1]{\endgroup\texttt{#1}}
\makeatother
\providecommand*\mcitethebibliography{\thebibliography}
\csname @ifundefined\endcsname{endmcitethebibliography}
  {\let\endmcitethebibliography\endthebibliography}{}



\begin{mcitethebibliography}{74}
\providecommand*\natexlab[1]{#1}
\providecommand*\mciteSetBstSublistMode[1]{}
\providecommand*\mciteSetBstMaxWidthForm[2]{}
\providecommand*\mciteBstWouldAddEndPuncttrue
  {\def\EndOfBibitem{\unskip.}}
\providecommand*\mciteBstWouldAddEndPunctfalse
  {\let\EndOfBibitem\relax}
\providecommand*\mciteSetBstMidEndSepPunct[3]{}
\providecommand*\mciteSetBstSublistLabelBeginEnd[3]{}
\providecommand*\EndOfBibitem{}
\mciteSetBstSublistMode{f}
\mciteSetBstMaxWidthForm{subitem}{(\alph{mcitesubitemcount})}
\mciteSetBstSublistLabelBeginEnd
  {\mcitemaxwidthsubitemform\space}
  {\relax}
  {\relax}

\bibitem[Banani \latin{et~al.}(2017)Banani, Lee, Hyman, and
  Rosen]{banani_biomolecular_2017}
Banani,~S.~F.; Lee,~H.~O.; Hyman,~A.~A.; Rosen,~M.~K. Biomolecular condensates:
  organizers of cellular biochemistry. \emph{Nature Reviews Molecular Cell
  Biology} \textbf{2017}, \emph{18}, 285--298, Number: 5 Publisher: Nature
  Publishing Group\relax
\mciteBstWouldAddEndPuncttrue
\mciteSetBstMidEndSepPunct{\mcitedefaultmidpunct}
{\mcitedefaultendpunct}{\mcitedefaultseppunct}\relax
\EndOfBibitem
\bibitem[Feric \latin{et~al.}(2016)Feric, Vaidya, Harmon, Mitrea, Zhu,
  Richardson, Kriwacki, Pappu, and Brangwynne]{feric_coexisting_2016}
Feric,~M.; Vaidya,~N.; Harmon,~T.~S.; Mitrea,~D.~M.; Zhu,~L.;
  Richardson,~T.~M.; Kriwacki,~R.~W.; Pappu,~R.~V.; Brangwynne,~C.~P.
  Coexisting {Liquid} {Phases} {Underlie} {Nucleolar} {Subcompartments}.
  \emph{Cell} \textbf{2016}, \emph{165}, 1686--1697, Publisher: Elsevier\relax
\mciteBstWouldAddEndPuncttrue
\mciteSetBstMidEndSepPunct{\mcitedefaultmidpunct}
{\mcitedefaultendpunct}{\mcitedefaultseppunct}\relax
\EndOfBibitem
\bibitem[Brangwynne \latin{et~al.}(2009)Brangwynne, Eckmann, Courson, Rybarska,
  Hoege, Gharakhani, J{\"U}licher, and Hyman]{brangwynne_germline_2009}
Brangwynne,~C.~P.; Eckmann,~C.~R.; Courson,~D.~S.; Rybarska,~A.; Hoege,~C.;
  Gharakhani,~J.; J{\"U}licher,~F.; Hyman,~A.~A. Germline {P} granules are liquid
  droplets that localize by controlled dissolution/condensation. \emph{Science
  (New York, N.Y.)} \textbf{2009}, \emph{324}, 1729--1732\relax
\mciteBstWouldAddEndPuncttrue
\mciteSetBstMidEndSepPunct{\mcitedefaultmidpunct}
{\mcitedefaultendpunct}{\mcitedefaultseppunct}\relax
\EndOfBibitem
\bibitem[Mittag and Pappu(2022)Mittag, and Pappu]{mittag_conceptual_2022}
Mittag,~T.; Pappu,~R.~V. A conceptual framework for understanding phase
  separation and addressing open questions and challenges. \emph{Molecular
  Cell} \textbf{2022}, \emph{82}, 2201--2214, Publisher: Elsevier\relax
\mciteBstWouldAddEndPuncttrue
\mciteSetBstMidEndSepPunct{\mcitedefaultmidpunct}
{\mcitedefaultendpunct}{\mcitedefaultseppunct}\relax
\EndOfBibitem
\bibitem[Babinchak \latin{et~al.}(2019)Babinchak, Haider, Dumm, Sarkar,
  Surewicz, Choi, and Surewicz]{babinchak_role_2019}
Babinchak,~W.~M.; Haider,~R.; Dumm,~B.~K.; Sarkar,~P.; Surewicz,~K.;
  Choi,~J.-K.; Surewicz,~W.~K. The role of liquid--liquid phase separation in
  aggregation of the {TDP}-43 low-complexity domain. \emph{Journal of
  Biological Chemistry} \textbf{2019}, \emph{294}, 6306--6317, Publisher:
  Elsevier\relax
\mciteBstWouldAddEndPuncttrue
\mciteSetBstMidEndSepPunct{\mcitedefaultmidpunct}
{\mcitedefaultendpunct}{\mcitedefaultseppunct}\relax
\EndOfBibitem
\bibitem[Alberti and Dormann(2019)Alberti, and
  Dormann]{alberti_liquidliquid_2019}
Alberti,~S.; Dormann,~D. Liquid--{Liquid} {Phase} {Separation} in {Disease}.
  \emph{Annual Review of Genetics} \textbf{2019}, \emph{53}, 171--194,
  \_eprint: https://doi.org/10.1146/annurev-genet-112618-043527\relax
\mciteBstWouldAddEndPuncttrue
\mciteSetBstMidEndSepPunct{\mcitedefaultmidpunct}
{\mcitedefaultendpunct}{\mcitedefaultseppunct}\relax
\EndOfBibitem
\bibitem[Do \latin{et~al.}(2022)Do, Lee, Lee, Kim, and
  Shin]{do_engineering_2022}
Do,~S.; Lee,~C.; Lee,~T.; Kim,~D.-N.; Shin,~Y. Engineering {DNA}-based
  synthetic condensates with programmable material properties, compositions,
  and functionalities. \emph{Science Advances} \textbf{2022}, \emph{8},
  eabj1771, Publisher: American Association for the Advancement of
  Science\relax
\mciteBstWouldAddEndPuncttrue
\mciteSetBstMidEndSepPunct{\mcitedefaultmidpunct}
{\mcitedefaultendpunct}{\mcitedefaultseppunct}\relax
\EndOfBibitem
\bibitem[Qian \latin{et~al.}(2022)Qian, Huang, and Xia]{qian_synthetic_2022}
Qian,~Z.-G.; Huang,~S.-C.; Xia,~X.-X. Synthetic protein condensates for
  cellular and metabolic engineering. \emph{Nature Chemical Biology}
  \textbf{2022}, \emph{18}, 1330--1340, Publisher: Nature Publishing
  Group\relax
\mciteBstWouldAddEndPuncttrue
\mciteSetBstMidEndSepPunct{\mcitedefaultmidpunct}
{\mcitedefaultendpunct}{\mcitedefaultseppunct}\relax
\EndOfBibitem
\bibitem[Oldfield and Dunker(2014)Oldfield, and
  Dunker]{oldfield_intrinsically_2014}
Oldfield,~C.~J.; Dunker,~A.~K. Intrinsically {Disordered} {Proteins} and
  {Intrinsically} {Disordered} {Protein} {Regions}. \emph{Annual Review of
  Biochemistry} \textbf{2014}, \emph{83}, 553--584, Publisher: Annual
  Reviews\relax
\mciteBstWouldAddEndPuncttrue
\mciteSetBstMidEndSepPunct{\mcitedefaultmidpunct}
{\mcitedefaultendpunct}{\mcitedefaultseppunct}\relax
\EndOfBibitem
\bibitem[Holehouse and Kragelund(2024)Holehouse, and
  Kragelund]{holehouse_molecular_2024}
Holehouse,~A.~S.; Kragelund,~B.~B. The molecular basis for cellular function of
  intrinsically disordered protein regions. \emph{Nature Reviews Molecular Cell
  Biology} \textbf{2024}, \emph{25}, 187--211, Publisher: Nature Publishing
  Group\relax
\mciteBstWouldAddEndPuncttrue
\mciteSetBstMidEndSepPunct{\mcitedefaultmidpunct}
{\mcitedefaultendpunct}{\mcitedefaultseppunct}\relax
\EndOfBibitem
\bibitem[Pappu \latin{et~al.}(2023)Pappu, Cohen, Dar, Farag, and
  Kar]{pappu_phase_2023}
Pappu,~R.~V.; Cohen,~S.~R.; Dar,~F.; Farag,~M.; Kar,~M. Phase {Transitions} of
  {Associative} {Biomacromolecules}. \emph{Chemical Reviews} \textbf{2023},
  \emph{123}, 8945--8987, Publisher: American Chemical Society\relax
\mciteBstWouldAddEndPuncttrue
\mciteSetBstMidEndSepPunct{\mcitedefaultmidpunct}
{\mcitedefaultendpunct}{\mcitedefaultseppunct}\relax
\EndOfBibitem
\bibitem[Nott \latin{et~al.}(2015)Nott, Petsalaki, Farber, Jervis, Fussner,
  Plochowietz, Craggs, Bazett-Jones, Pawson, Forman-Kay, and
  Baldwin]{nott_phase_2015}
Nott,~T.~J.; Petsalaki,~E.; Farber,~P.; Jervis,~D.; Fussner,~E.;
  Plochowietz,~A.; Craggs,~T.~D.; Bazett-Jones,~D.~P.; Pawson,~T.;
  Forman-Kay,~J.~D. \latin{et~al.}  Phase {Transition} of a {Disordered}
  {Nuage} {Protein} {Generates} {Environmentally} {Responsive} {Membraneless}
  {Organelles}. \emph{Molecular Cell} \textbf{2015}, \emph{57}, 936--947\relax
\mciteBstWouldAddEndPuncttrue
\mciteSetBstMidEndSepPunct{\mcitedefaultmidpunct}
{\mcitedefaultendpunct}{\mcitedefaultseppunct}\relax
\EndOfBibitem
\bibitem[Wang \latin{et~al.}(2018)Wang, Choi, Holehouse, Lee, Zhang, Jahnel,
  Maharana, Lemaitre, Pozniakovsky, Drechsel, Poser, Pappu, Alberti, and
  Hyman]{wang_molecular_2018}
Wang,~J.; Choi,~J.-M.; Holehouse,~A.~S.; Lee,~H.~O.; Zhang,~X.; Jahnel,~M.;
  Maharana,~S.; Lemaitre,~R.; Pozniakovsky,~A.; Drechsel,~D. \latin{et~al.}  A
  {Molecular} {Grammar} {Governing} the {Driving} {Forces} for {Phase}
  {Separation} of {Prion}-like {RNA} {Binding} {Proteins}. \emph{Cell}
  \textbf{2018}, \emph{174}, 688--699.e16\relax
\mciteBstWouldAddEndPuncttrue
\mciteSetBstMidEndSepPunct{\mcitedefaultmidpunct}
{\mcitedefaultendpunct}{\mcitedefaultseppunct}\relax
\EndOfBibitem
\bibitem[Schuster \latin{et~al.}(2020)Schuster, Dignon, Tang, Kelley,
  Ranganath, Jahnke, Simpkins, Regy, Hammer, Good, and
  Mittal]{schuster_identifying_2020}
Schuster,~B.~S.; Dignon,~G.~L.; Tang,~W.~S.; Kelley,~F.~M.; Ranganath,~A.~K.;
  Jahnke,~C.~N.; Simpkins,~A.~G.; Regy,~R.~M.; Hammer,~D.~A.; Good,~M.~C.
  \latin{et~al.}  Identifying sequence perturbations to an intrinsically
  disordered protein that determine its phase-separation behavior.
  \emph{Proceedings of the National Academy of Sciences} \textbf{2020},
  \emph{117}, 11421--11431, Publisher: Proceedings of the National Academy of
  Sciences\relax
\mciteBstWouldAddEndPuncttrue
\mciteSetBstMidEndSepPunct{\mcitedefaultmidpunct}
{\mcitedefaultendpunct}{\mcitedefaultseppunct}\relax
\EndOfBibitem
\bibitem[Brangwynne \latin{et~al.}(2015)Brangwynne, Tompa, and
  Pappu]{brangwynne_polymer_2015}
Brangwynne,~C.~P.; Tompa,~P.; Pappu,~R.~V. Polymer physics of intracellular
  phase transitions. \emph{Nature Physics} \textbf{2015}, \emph{11}, 899--904,
  Number: 11 Publisher: Nature Publishing Group\relax
\mciteBstWouldAddEndPuncttrue
\mciteSetBstMidEndSepPunct{\mcitedefaultmidpunct}
{\mcitedefaultendpunct}{\mcitedefaultseppunct}\relax
\EndOfBibitem
\bibitem[Lin and Chan(2017)Lin, and Chan]{lin_phase_2017}
Lin,~Y.-H.; Chan,~H.~S. Phase {Separation} and {Single}-{Chain} {Compactness}
  of {Charged} {Disordered} {Proteins} {Are} {Strongly} {Correlated}.
  \emph{Biophysical Journal} \textbf{2017}, \emph{112}, 2043--2046\relax
\mciteBstWouldAddEndPuncttrue
\mciteSetBstMidEndSepPunct{\mcitedefaultmidpunct}
{\mcitedefaultendpunct}{\mcitedefaultseppunct}\relax
\EndOfBibitem
\bibitem[Statt \latin{et~al.}(2020)Statt, Casademunt, Brangwynne, and
  Panagiotopoulos]{statt_model_2020}
Statt,~A.; Casademunt,~H.; Brangwynne,~C.~P.; Panagiotopoulos,~A.~Z. Model for
  disordered proteins with strongly sequence-dependent liquid phase behavior.
  \emph{The Journal of Chemical Physics} \textbf{2020}, \emph{152},
  075101\relax
\mciteBstWouldAddEndPuncttrue
\mciteSetBstMidEndSepPunct{\mcitedefaultmidpunct}
{\mcitedefaultendpunct}{\mcitedefaultseppunct}\relax
\EndOfBibitem
\bibitem[Rekhi \latin{et~al.}(2023)Rekhi, Sundaravadivelu~Devarajan, Howard,
  Kim, Nikoubashman, and Mittal]{rekhi_role_2023}
Rekhi,~S.; Sundaravadivelu~Devarajan,~D.; Howard,~M.~P.; Kim,~Y.~C.;
  Nikoubashman,~A.; Mittal,~J. Role of {Strong} {Localized} vs {Weak}
  {Distributed} {Interactions} in {Disordered} {Protein} {Phase} {Separation}.
  \emph{The Journal of Physical Chemistry B} \textbf{2023}, \emph{127},
  3829--3838, Publisher: American Chemical Society\relax
\mciteBstWouldAddEndPuncttrue
\mciteSetBstMidEndSepPunct{\mcitedefaultmidpunct}
{\mcitedefaultendpunct}{\mcitedefaultseppunct}\relax
\EndOfBibitem
\bibitem[Das and Pappu(2013)Das, and Pappu]{das_conformations_2013}
Das,~R.~K.; Pappu,~R.~V. Conformations of intrinsically disordered proteins are
  influenced by linear sequence distributions of oppositely charged residues.
  \emph{Proceedings of the National Academy of Sciences} \textbf{2013},
  \emph{110}, 13392--13397, Publisher: Proceedings of the National Academy of
  Sciences\relax
\mciteBstWouldAddEndPuncttrue
\mciteSetBstMidEndSepPunct{\mcitedefaultmidpunct}
{\mcitedefaultendpunct}{\mcitedefaultseppunct}\relax
\EndOfBibitem
\bibitem[Regy \latin{et~al.}(2021)Regy, Thompson, Kim, and
  Mittal]{regy_improved_2021}
Regy,~R.~M.; Thompson,~J.; Kim,~Y.~C.; Mittal,~J. Improved coarse-grained model
  for studying sequence dependent phase separation of disordered proteins.
  \emph{Protein Science} \textbf{2021}, \emph{30}, 1371--1379, \_eprint:
  https://onlinelibrary.wiley.com/doi/pdf/10.1002/pro.4094\relax
\mciteBstWouldAddEndPuncttrue
\mciteSetBstMidEndSepPunct{\mcitedefaultmidpunct}
{\mcitedefaultendpunct}{\mcitedefaultseppunct}\relax
\EndOfBibitem
\bibitem[Joseph \latin{et~al.}(2022)Joseph, Reinhardt, Aguirre, Yu~Chew,
  Russell, Espinosa, Garaizar, and
  Collepardo-Guevara]{joseph_physics-driven_2022}
Joseph,~J.~A.; Reinhardt,~A.; Aguirre,~A.; Yu~Chew,~P.; Russell,~K.~O.;
  Espinosa,~J.~R.; Garaizar,~A.; Collepardo-Guevara,~R. Physics-driven
  coarse-grained model for biomolecular phase separation with near-quantitative
  accuracy. \emph{Biophysical Journal} \textbf{2022}, \emph{121}, 307a\relax
\mciteBstWouldAddEndPuncttrue
\mciteSetBstMidEndSepPunct{\mcitedefaultmidpunct}
{\mcitedefaultendpunct}{\mcitedefaultseppunct}\relax
\EndOfBibitem
\bibitem[Tesei \latin{et~al.}(2021)Tesei, Schulze, Crehuet, and
  Lindorff-Larsen]{tesei_accurate_2021}
Tesei,~G.; Schulze,~T.~K.; Crehuet,~R.; Lindorff-Larsen,~K. Accurate model of
  liquid-liquid phase behavior of intrinsically disordered proteins from
  optimization of single-chain properties. \emph{Proceedings of the National
  Academy of Sciences of the United States of America} \textbf{2021},
  \emph{118}, e2111696118\relax
\mciteBstWouldAddEndPuncttrue
\mciteSetBstMidEndSepPunct{\mcitedefaultmidpunct}
{\mcitedefaultendpunct}{\mcitedefaultseppunct}\relax
\EndOfBibitem
\bibitem[Tesei and Lindorff-Larsen(2022)Tesei, and
  Lindorff-Larsen]{tesei_improved_2022}
Tesei,~G.; Lindorff-Larsen,~K. Improved predictions of phase behaviour of
  intrinsically disordered proteins by tuning the interaction range. \emph{Open
  Research Europe} \textbf{2022}, \emph{2}, 94\relax
\mciteBstWouldAddEndPuncttrue
\mciteSetBstMidEndSepPunct{\mcitedefaultmidpunct}
{\mcitedefaultendpunct}{\mcitedefaultseppunct}\relax
\EndOfBibitem
\bibitem[Li \latin{et~al.}(2020)Li, Peng, Li, Tang, Zhu, Huang, Qi, and
  Zhang]{li_llpsdb_2020}
Li,~Q.; Peng,~X.; Li,~Y.; Tang,~W.; Zhu,~J.; Huang,~J.; Qi,~Y.; Zhang,~Z.
  {LLPSDB}: a database of proteins undergoing liquid--liquid phase separation
  in vitro. \emph{Nucleic Acids Research} \textbf{2020}, \emph{48},
  D320--D327\relax
\mciteBstWouldAddEndPuncttrue
\mciteSetBstMidEndSepPunct{\mcitedefaultmidpunct}
{\mcitedefaultendpunct}{\mcitedefaultseppunct}\relax
\EndOfBibitem
\bibitem[M{\'e}sz{\'a}ros \latin{et~al.}(2020)M{\'e}sz{\' a}ros, Erd{\H o}s, Szab{\'o}, Sch{\'a}d,
  Tantos, Abukhairan, Horv{\'a}th, Murvai, Kov{\'a}cs, Kov{\'a}cs, Tosatto, Tompa,
  Doszt{\'a}nyi, and Pancsa]{meszaros_phasepro_2020}
M{\'e}sz{\'a}ros,~B.; Erd{\H o}s,~G.; Szab{\'o},~B.; Sch{\'a}d,~{\'E}.; Tantos,~{\'A}.; Abukhairan,~R.;
  Horv{\'a}th,~T.; Murvai,~N.; Kov{\'a}cs,~O.~P.; Kov{\'a}cs,~M. \latin{et~al.}
  {PhaSePro}: the database of proteins driving liquid--liquid phase
  separation. \emph{Nucleic Acids Research} \textbf{2020}, \emph{48},
  D360--D367\relax
\mciteBstWouldAddEndPuncttrue
\mciteSetBstMidEndSepPunct{\mcitedefaultmidpunct}
{\mcitedefaultendpunct}{\mcitedefaultseppunct}\relax
\EndOfBibitem
\bibitem[van Mierlo \latin{et~al.}(2021)van Mierlo, Jansen, Wang, Poser, van
  Heeringen, and Vermeulen]{van_mierlo_predicting_2021}
van Mierlo,~G.; Jansen,~J. R.~G.; Wang,~J.; Poser,~I.; van Heeringen,~S.~J.;
  Vermeulen,~M. Predicting protein condensate formation using machine learning.
  \emph{Cell Reports} \textbf{2021}, \emph{34}, 108705\relax
\mciteBstWouldAddEndPuncttrue
\mciteSetBstMidEndSepPunct{\mcitedefaultmidpunct}
{\mcitedefaultendpunct}{\mcitedefaultseppunct}\relax
\EndOfBibitem
\bibitem[Chu \latin{et~al.}(2022)Chu, Sun, Li, Xu, Zhang, Lai, and
  Pei]{chu_prediction_2022}
Chu,~X.; Sun,~T.; Li,~Q.; Xu,~Y.; Zhang,~Z.; Lai,~L.; Pei,~J. Prediction of
  liquid--liquid phase separating proteins using machine learning. \emph{BMC
  Bioinformatics} \textbf{2022}, \emph{23}, 72\relax
\mciteBstWouldAddEndPuncttrue
\mciteSetBstMidEndSepPunct{\mcitedefaultmidpunct}
{\mcitedefaultendpunct}{\mcitedefaultseppunct}\relax
\EndOfBibitem
\bibitem[Chen \latin{et~al.}(2022)Chen, Hou, Wang, Yu, Chen, Shen, Hou, Li, and
  Li]{chen_screening_2022}
Chen,~Z.; Hou,~C.; Wang,~L.; Yu,~C.; Chen,~T.; Shen,~B.; Hou,~Y.; Li,~P.;
  Li,~T. Screening membraneless organelle participants with machine-learning
  models that integrate multimodal features. \emph{Proceedings of the National
  Academy of Sciences} \textbf{2022}, \emph{119}, e2115369119, Publisher:
  Proceedings of the National Academy of Sciences\relax
\mciteBstWouldAddEndPuncttrue
\mciteSetBstMidEndSepPunct{\mcitedefaultmidpunct}
{\mcitedefaultendpunct}{\mcitedefaultseppunct}\relax
\EndOfBibitem
\bibitem[Sun \latin{et~al.}(2024)Sun, Qu, Zhao, Zhang, Liu, Wang, Wei, Liu,
  Wang, Zeng, Tang, Ling, Qing, Jiang, Chen, Chen, Kuang, Gao, Zeng, Huang,
  Yuan, Fan, Yu, and Ding]{sun_precise_2024}
Sun,~J.; Qu,~J.; Zhao,~C.; Zhang,~X.; Liu,~X.; Wang,~J.; Wei,~C.; Liu,~X.;
  Wang,~M.; Zeng,~P. \latin{et~al.}  Precise prediction of phase-separation key
  residues by machine learning. \emph{Nature Communications} \textbf{2024},
  \emph{15}, 2662, Publisher: Nature Publishing Group\relax
\mciteBstWouldAddEndPuncttrue
\mciteSetBstMidEndSepPunct{\mcitedefaultmidpunct}
{\mcitedefaultendpunct}{\mcitedefaultseppunct}\relax
\EndOfBibitem
\bibitem[Tesei \latin{et~al.}(2024)Tesei, Trolle, Jonsson, Betz, Knudsen,
  Pesce, Johansson, and Lindorff-Larsen]{tesei_conformational_2024}
Tesei,~G.; Trolle,~A.~I.; Jonsson,~N.; Betz,~J.; Knudsen,~F.~E.; Pesce,~F.;
  Johansson,~K.~E.; Lindorff-Larsen,~K. Conformational ensembles of the human
  intrinsically disordered proteome. \emph{Nature} \textbf{2024}, 1--8,
  Publisher: Nature Publishing Group\relax
\mciteBstWouldAddEndPuncttrue
\mciteSetBstMidEndSepPunct{\mcitedefaultmidpunct}
{\mcitedefaultendpunct}{\mcitedefaultseppunct}\relax
\EndOfBibitem
\bibitem[An \latin{et~al.}(2024)An, Webb, and Jacobs]{an_active_2024}
An,~Y.; Webb,~M.~A.; Jacobs,~W.~M. Active learning of the
  thermodynamics-dynamics trade-off in protein condensates. \emph{Science
  Advances} \textbf{2024}, \emph{10}, eadj2448, Publisher: American Association
  for the Advancement of Science\relax
\mciteBstWouldAddEndPuncttrue
\mciteSetBstMidEndSepPunct{\mcitedefaultmidpunct}
{\mcitedefaultendpunct}{\mcitedefaultseppunct}\relax
\EndOfBibitem
\bibitem[von B{\"u}low \latin{et~al.}(2025)von B{\"u}low, Tesei, Zaidi, Mittag, and
  Lindorff-Larsen]{von_bulow_prediction_2025}
von B{\"u}low,~S.; Tesei,~G.; Zaidi,~F.~K.; Mittag,~T.; Lindorff-Larsen,~K.
  Prediction of phase-separation propensities of disordered proteins from
  sequence. \emph{Proceedings of the National Academy of Sciences}
  \textbf{2025}, \emph{122}, e2417920122, Publisher: Proceedings of the
  National Academy of Sciences\relax
\mciteBstWouldAddEndPuncttrue
\mciteSetBstMidEndSepPunct{\mcitedefaultmidpunct}
{\mcitedefaultendpunct}{\mcitedefaultseppunct}\relax
\EndOfBibitem
\bibitem[Changiarath \latin{et~al.}(2025)Changiarath, Arya, Xenidis, Padeken,
  and Stelzl]{changiarath_sequence_2025}
Changiarath,~A.; Arya,~A.; Xenidis,~V.~A.; Padeken,~J.; Stelzl,~L.~S. Sequence
  determinants of protein phase separation and recognition by protein
  phase-separated condensates through molecular dynamics and active learning.
  \emph{Faraday Discussions} \textbf{2025}, \emph{256}, 235--254, Publisher:
  The Royal Society of Chemistry\relax
\mciteBstWouldAddEndPuncttrue
\mciteSetBstMidEndSepPunct{\mcitedefaultmidpunct}
{\mcitedefaultendpunct}{\mcitedefaultseppunct}\relax
\EndOfBibitem
\bibitem[Lotthammer \latin{et~al.}(2024)Lotthammer, Ginell, Griffith,
  Emenecker, and Holehouse]{lotthammer_direct_2024}
Lotthammer,~J.~M.; Ginell,~G.~M.; Griffith,~D.; Emenecker,~R.~J.;
  Holehouse,~A.~S. Direct prediction of intrinsically disordered protein
  conformational properties from sequences. \emph{Nature Methods}
  \textbf{2024}, 1--12, Publisher: Nature Publishing Group\relax
\mciteBstWouldAddEndPuncttrue
\mciteSetBstMidEndSepPunct{\mcitedefaultmidpunct}
{\mcitedefaultendpunct}{\mcitedefaultseppunct}\relax
\EndOfBibitem
\bibitem[Pesce \latin{et~al.}(2024)Pesce, Bremer, Tesei, Hopkins, Grace,
  Mittag, and Lindorff-Larsen]{pesce_design_2024}
Pesce,~F.; Bremer,~A.; Tesei,~G.; Hopkins,~J.~B.; Grace,~C.~R.; Mittag,~T.;
  Lindorff-Larsen,~K. Design of intrinsically disordered protein variants with
  diverse structural properties. \emph{Science Advances} \textbf{2024},
  \emph{10}, eadm9926, Publisher: American Association for the Advancement of
  Science\relax
\mciteBstWouldAddEndPuncttrue
\mciteSetBstMidEndSepPunct{\mcitedefaultmidpunct}
{\mcitedefaultendpunct}{\mcitedefaultseppunct}\relax
\EndOfBibitem
\bibitem[Dignon \latin{et~al.}(2018)Dignon, Zheng, Kim, Best, and
  Mittal]{dignon_sequence_2018}
Dignon,~G.~L.; Zheng,~W.; Kim,~Y.~C.; Best,~R.~B.; Mittal,~J. Sequence
  determinants of protein phase behavior from a coarse-grained model.
  \emph{PLOS Computational Biology} \textbf{2018}, \emph{14}, e1005941,
  Publisher: Public Library of Science\relax
\mciteBstWouldAddEndPuncttrue
\mciteSetBstMidEndSepPunct{\mcitedefaultmidpunct}
{\mcitedefaultendpunct}{\mcitedefaultseppunct}\relax
\EndOfBibitem
\bibitem[Zeng \latin{et~al.}(2020)Zeng, Holehouse, Chilkoti, Mittag, and
  Pappu]{zeng_connecting_2020}
Zeng,~X.; Holehouse,~A.~S.; Chilkoti,~A.; Mittag,~T.; Pappu,~R.~V. Connecting
  {Coil}-to-{Globule} {Transitions} to {Full} {Phase} {Diagrams} for
  {Intrinsically} {Disordered} {Proteins}. \emph{Biophysical Journal}
  \textbf{2020}, \emph{119}, 402--418\relax
\mciteBstWouldAddEndPuncttrue
\mciteSetBstMidEndSepPunct{\mcitedefaultmidpunct}
{\mcitedefaultendpunct}{\mcitedefaultseppunct}\relax
\EndOfBibitem
\bibitem[Dignon \latin{et~al.}(2018)Dignon, Zheng, Best, Kim, and
  Mittal]{dignon_relation_2018}
Dignon,~G.~L.; Zheng,~W.; Best,~R.~B.; Kim,~Y.~C.; Mittal,~J. Relation between
  single-molecule properties and phase behavior of intrinsically disordered
  proteins. \emph{Proceedings of the National Academy of Sciences}
  \textbf{2018}, \emph{115}, 9929--9934, Publisher: Proceedings of the National
  Academy of Sciences\relax
\mciteBstWouldAddEndPuncttrue
\mciteSetBstMidEndSepPunct{\mcitedefaultmidpunct}
{\mcitedefaultendpunct}{\mcitedefaultseppunct}\relax
\EndOfBibitem
\bibitem[Zeng \latin{et~al.}(2021)Zeng, Liu, Fossat, Ren, Chilkoti, and
  Pappu]{zeng_design_2021}
Zeng,~X.; Liu,~C.; Fossat,~M.~J.; Ren,~P.; Chilkoti,~A.; Pappu,~R.~V. Design of
  intrinsically disordered proteins that undergo phase transitions with lower
  critical solution temperatures. \emph{APL Materials} \textbf{2021}, \emph{9},
  021119\relax
\mciteBstWouldAddEndPuncttrue
\mciteSetBstMidEndSepPunct{\mcitedefaultmidpunct}
{\mcitedefaultendpunct}{\mcitedefaultseppunct}\relax
\EndOfBibitem
\bibitem[Quigley and Williams(2015)Quigley, and Williams]{quigley_second_2015}
Quigley,~A.; Williams,~D.~R. The second virial coefficient as a predictor of
  protein aggregation propensity: {A} self-interaction chromatography study.
  \emph{European Journal of Pharmaceutics and Biopharmaceutics} \textbf{2015},
  \emph{96}, 282--290\relax
\mciteBstWouldAddEndPuncttrue
\mciteSetBstMidEndSepPunct{\mcitedefaultmidpunct}
{\mcitedefaultendpunct}{\mcitedefaultseppunct}\relax
\EndOfBibitem
\bibitem[Adachi and Kawaguchi(2024)Adachi, and
  Kawaguchi]{adachi_predicting_2024}
Adachi,~K.; Kawaguchi,~K. Predicting {Heteropolymer} {Interactions}: {Demixing}
  and {Hypermixing} of {Disordered} {Protein} {Sequences}. \emph{Physical
  Review X} \textbf{2024}, \emph{14}, 031011, Publisher: American Physical
  Society\relax
\mciteBstWouldAddEndPuncttrue
\mciteSetBstMidEndSepPunct{\mcitedefaultmidpunct}
{\mcitedefaultendpunct}{\mcitedefaultseppunct}\relax
\EndOfBibitem
\bibitem[Dignon \latin{et~al.}(2019)Dignon, Zheng, Kim, and
  Mittal]{dignon_temperature-controlled_2019}
Dignon,~G.~L.; Zheng,~W.; Kim,~Y.~C.; Mittal,~J. Temperature-{Controlled}
  {Liquid}--{Liquid} {Phase} {Separation} of {Disordered} {Proteins}.
  \emph{ACS Central Science} \textbf{2019}, \emph{5}, 821--830, Publisher:
  American Chemical Society\relax
\mciteBstWouldAddEndPuncttrue
\mciteSetBstMidEndSepPunct{\mcitedefaultmidpunct}
{\mcitedefaultendpunct}{\mcitedefaultseppunct}\relax
\EndOfBibitem
\bibitem[Pal \latin{et~al.}(2024)Pal, Wess{\'e}n, Das, and
  Chan]{pal_differential_2024}
Pal,~T.; Wess{\'e}n,~J.; Das,~S.; Chan,~H.~S. Differential {Effects} of
  {Sequence}-{Local} versus {Nonlocal} {Charge} {Patterns} on {Phase}
  {Separation} and {Conformational} {Dimensions} of {Polyampholytes} as {Model}
  {Intrinsically} {Disordered} {Proteins}. \emph{The Journal of Physical
  Chemistry Letters} \textbf{2024}, \emph{15}, 8248--8256, Publisher: American
  Chemical Society\relax
\mciteBstWouldAddEndPuncttrue
\mciteSetBstMidEndSepPunct{\mcitedefaultmidpunct}
{\mcitedefaultendpunct}{\mcitedefaultseppunct}\relax
\EndOfBibitem
\bibitem[Jin \latin{et~al.}(2025)Jin, Oliver, Webb, and
  Jacobs]{jin_predicting_2025}
Jin,~J.; Oliver,~W.; Webb,~M.~A.; Jacobs,~W.~M. Predicting heteropolymer phase
  separation using two-chain contact maps. \emph{The Journal of Chemical
  Physics} \textbf{2025}, \emph{163}, 014102\relax
\mciteBstWouldAddEndPuncttrue
\mciteSetBstMidEndSepPunct{\mcitedefaultmidpunct}
{\mcitedefaultendpunct}{\mcitedefaultseppunct}\relax
\EndOfBibitem
\bibitem[Webb \latin{et~al.}(2023)Webb, Jacobs, An, and
  Oliver]{webb_thermodynamic_2023}
Webb,~M.; Jacobs,~W.; An,~Y.; Oliver,~W. Thermodynamic and dynamics data for
  coarse-grained intrinsically disordered proteins generated by active
  learning. \textbf{2023}, Publisher: Princeton University\relax
\mciteBstWouldAddEndPuncttrue
\mciteSetBstMidEndSepPunct{\mcitedefaultmidpunct}
{\mcitedefaultendpunct}{\mcitedefaultseppunct}\relax
\EndOfBibitem
\bibitem[Sickmeier \latin{et~al.}(2007)Sickmeier, Hamilton, LeGall, Vacic,
  Cortese, Tantos, Szabo, Tompa, Chen, Uversky, Obradovic, and
  Dunker]{sickmeier_disprot_2007}
Sickmeier,~M.; Hamilton,~J.~A.; LeGall,~T.; Vacic,~V.; Cortese,~M.~S.;
  Tantos,~A.; Szabo,~B.; Tompa,~P.; Chen,~J.; Uversky,~V.~N. \latin{et~al.}
  {DisProt}: the {Database} of {Disordered} {Proteins}. \emph{Nucleic Acids
  Research} \textbf{2007}, \emph{35}, D786--D793\relax
\mciteBstWouldAddEndPuncttrue
\mciteSetBstMidEndSepPunct{\mcitedefaultmidpunct}
{\mcitedefaultendpunct}{\mcitedefaultseppunct}\relax
\EndOfBibitem
\bibitem[Urry \latin{et~al.}(1992)Urry, Gowda, Parker, Luan, Reid, Harris,
  Pattanaik, and Harris]{urry_hydrophobicity_1992}
Urry,~D.~W.; Gowda,~D.~C.; Parker,~T.~M.; Luan,~C.~H.; Reid,~M.~C.;
  Harris,~C.~M.; Pattanaik,~A.; Harris,~R.~D. Hydrophobicity scale for proteins
  based on inverse temperature transitions. \emph{Biopolymers} \textbf{1992},
  \emph{32}, 1243--1250\relax
\mciteBstWouldAddEndPuncttrue
\mciteSetBstMidEndSepPunct{\mcitedefaultmidpunct}
{\mcitedefaultendpunct}{\mcitedefaultseppunct}\relax
\EndOfBibitem
\bibitem[Thompson \latin{et~al.}(2022)Thompson, Aktulga, Berger, Bolintineanu,
  Brown, Crozier, in~'t Veld, Kohlmeyer, Moore, Nguyen, Shan, Stevens,
  Tranchida, Trott, and Plimpton]{thompson_lammps_2022}
Thompson,~A.~P.; Aktulga,~H.~M.; Berger,~R.; Bolintineanu,~D.~S.; Brown,~W.~M.;
  Crozier,~P.~S.; in~'t Veld,~P.~J.; Kohlmeyer,~A.; Moore,~S.~G.; Nguyen,~T.~D.
  \latin{et~al.}  {LAMMPS} - a flexible simulation tool for particle-based
  materials modeling at the atomic, meso, and continuum scales. \emph{Computer
  Physics Communications} \textbf{2022}, \emph{271}, 108171\relax
\mciteBstWouldAddEndPuncttrue
\mciteSetBstMidEndSepPunct{\mcitedefaultmidpunct}
{\mcitedefaultendpunct}{\mcitedefaultseppunct}\relax
\EndOfBibitem
\bibitem[Binder \latin{et~al.}(2012)Binder, Block, Virnau, and
  Tr{\"o}ster]{binder_beyond_2012}
Binder,~K.; Block,~B.~J.; Virnau,~P.; Tr{\"o}ster,~A. Beyond the {Van} {Der}
  {Waals} loop: {What} can be learned from simulating {Lennard}-{Jones} fluids
  inside the region of phase coexistence. \emph{American Journal of Physics}
  \textbf{2012}, \emph{80}, 1099--1109\relax
\mciteBstWouldAddEndPuncttrue
\mciteSetBstMidEndSepPunct{\mcitedefaultmidpunct}
{\mcitedefaultendpunct}{\mcitedefaultseppunct}\relax
\EndOfBibitem
\bibitem[Virtanen \latin{et~al.}(2020)Virtanen, Gommers, Oliphant, Haberland,
  Reddy, Cournapeau, Burovski, Peterson, Weckesser, Bright, van~der Walt,
  Brett, Wilson, Millman, Mayorov, Nelson, Jones, Kern, Larson, Carey, Polat,
  Feng, Moore, VanderPlas, Laxalde, Perktold, Cimrman, Henriksen, Quintero,
  Harris, Archibald, Ribeiro, Pedregosa, and van Mulbregt]{virtanen_scipy_2020}
Virtanen,~P.; Gommers,~R.; Oliphant,~T.~E.; Haberland,~M.; Reddy,~T.;
  Cournapeau,~D.; Burovski,~E.; Peterson,~P.; Weckesser,~W.; Bright,~J.
  \latin{et~al.}  {SciPy} 1.0: fundamental algorithms for scientific computing
  in {Python}. \emph{Nature Methods} \textbf{2020}, \emph{17}, 261--272,
  Publisher: Nature Publishing Group\relax
\mciteBstWouldAddEndPuncttrue
\mciteSetBstMidEndSepPunct{\mcitedefaultmidpunct}
{\mcitedefaultendpunct}{\mcitedefaultseppunct}\relax
\EndOfBibitem
\bibitem[Wang and Wang(2014)Wang, and Wang]{R:2014_Wang_Theory}
Wang,~R.; Wang,~Z.-G. Theory of Polymer Chains in Poor Solvent: Single-Chain
  Structure, Solution Thermodynamics, and Œò Point. \emph{Macromolecules}
  \textbf{2014}, \emph{47}, 4094--4102\relax
\mciteBstWouldAddEndPuncttrue
\mciteSetBstMidEndSepPunct{\mcitedefaultmidpunct}
{\mcitedefaultendpunct}{\mcitedefaultseppunct}\relax
\EndOfBibitem
\bibitem[Xu and Wang(2021)Xu, and Wang]{R:2021_Xu_Coil}
Xu,~Y.; Wang,~Z.-G. Coil-to-Globule Transition in Polymeric Solvents.
  \emph{Macromolecules} \textbf{2021}, \emph{54}, 10984--10993\relax
\mciteBstWouldAddEndPuncttrue
\mciteSetBstMidEndSepPunct{\mcitedefaultmidpunct}
{\mcitedefaultendpunct}{\mcitedefaultseppunct}\relax
\EndOfBibitem
\bibitem[Dhamankar and Webb(2024)Dhamankar, and
  Webb]{R:2024_Dhamankar_Asymmetry}
Dhamankar,~S.; Webb,~M.~A. Asymmetry in Polymer--Solvent Interactions Yields
  Complex Thermoresponsive Behavior. \emph{ACS Macro Letters} \textbf{2024},
  \emph{13}, 818--825\relax
\mciteBstWouldAddEndPuncttrue
\mciteSetBstMidEndSepPunct{\mcitedefaultmidpunct}
{\mcitedefaultendpunct}{\mcitedefaultseppunct}\relax
\EndOfBibitem
\bibitem[Rubinstein and Colby(2003)Rubinstein, and
  Colby]{rubinstein_polymer_2003}
Rubinstein,~M.; Colby,~R.~H. \emph{Polymer physics}; Oxford university press:
  Oxford New York, 2003\relax
\mciteBstWouldAddEndPuncttrue
\mciteSetBstMidEndSepPunct{\mcitedefaultmidpunct}
{\mcitedefaultendpunct}{\mcitedefaultseppunct}\relax
\EndOfBibitem
\bibitem[Raos and Allegra(1996)Raos, and Allegra]{raos_chain_1996}
Raos,~G.; Allegra,~G. Chain collapse and phase separation in poor-solvent
  polymer solutions: {A} unified molecular description. \emph{The Journal of
  Chemical Physics} \textbf{1996}, \emph{104}, 1626--1645\relax
\mciteBstWouldAddEndPuncttrue
\mciteSetBstMidEndSepPunct{\mcitedefaultmidpunct}
{\mcitedefaultendpunct}{\mcitedefaultseppunct}\relax
\EndOfBibitem
\bibitem[Comer \latin{et~al.}(2015)Comer, Gumbart, H{\'e}nin, Leli{\`e}vre,
  Pohorille, and Chipot]{comer_adaptive_2015}
Comer,~J.; Gumbart,~J.~C.; H{\'e}nin,~J.; Leli{\`e}vre,~T.; Pohorille,~A.; Chipot,~C.
  The {Adaptive} {Biasing} {Force} {Method}: {Everything} {You} {Always}
  {Wanted} {To} {Know} but {Were} {Afraid} {To} {Ask}. \emph{The Journal of
  Physical Chemistry B} \textbf{2015}, \emph{119}, 1129--1151, Publisher:
  American Chemical Society\relax
\mciteBstWouldAddEndPuncttrue
\mciteSetBstMidEndSepPunct{\mcitedefaultmidpunct}
{\mcitedefaultendpunct}{\mcitedefaultseppunct}\relax
\EndOfBibitem
\bibitem[Fiorin \latin{et~al.}(2013)Fiorin, Klein, and
  H{\'e}nin]{fiorin_using_2013}
Fiorin,~G.; Klein,~M.~L.; H{\'e}nin,~J. Using collective variables to drive
  molecular dynamics simulations. \emph{Molecular Physics} \textbf{2013},
  \emph{111}, 3345--3362, Publisher: Taylor \& Francis \_eprint:
  https://doi.org/10.1080/00268976.2013.813594\relax
\mciteBstWouldAddEndPuncttrue
\mciteSetBstMidEndSepPunct{\mcitedefaultmidpunct}
{\mcitedefaultendpunct}{\mcitedefaultseppunct}\relax
\EndOfBibitem
\bibitem[Patel and Webb(2023)Patel, and Webb]{patel_data-driven_2023}
Patel,~R.~A.; Webb,~M.~A. Data-{Driven} {Design} of {Polymer}-{Based}
  {Biomaterials}: {High}-throughput {Simulation}, {Experimentation}, and
  {Machine} {Learning}. \emph{ACS Applied Bio Materials} \textbf{2023},
  Publisher: American Chemical Society\relax
\mciteBstWouldAddEndPuncttrue
\mciteSetBstMidEndSepPunct{\mcitedefaultmidpunct}
{\mcitedefaultendpunct}{\mcitedefaultseppunct}\relax
\EndOfBibitem
\bibitem[Firman and Ghosh(2018)Firman, and Ghosh]{firman_sequence_2018}
Firman,~T.; Ghosh,~K. Sequence charge decoration dictates coil-globule
  transition in intrinsically disordered proteins. \emph{The Journal of
  Chemical Physics} \textbf{2018}, \emph{148}, 123305\relax
\mciteBstWouldAddEndPuncttrue
\mciteSetBstMidEndSepPunct{\mcitedefaultmidpunct}
{\mcitedefaultendpunct}{\mcitedefaultseppunct}\relax
\EndOfBibitem
\bibitem[Zheng \latin{et~al.}(2020)Zheng, Dignon, Brown, Kim, and
  Mittal]{zheng_hydropathy_2020}
Zheng,~W.; Dignon,~G.; Brown,~M.; Kim,~Y.~C.; Mittal,~J. Hydropathy
  {Patterning} {Complements} {Charge} {Patterning} to {Describe}
  {Conformational} {Preferences} of {Disordered} {Proteins}. \emph{The Journal
  of Physical Chemistry Letters} \textbf{2020}, \emph{11}, 3408--3415,
  Publisher: American Chemical Society\relax
\mciteBstWouldAddEndPuncttrue
\mciteSetBstMidEndSepPunct{\mcitedefaultmidpunct}
{\mcitedefaultendpunct}{\mcitedefaultseppunct}\relax
\EndOfBibitem
\bibitem[McInnes \latin{et~al.}(2020)McInnes, Healy, and
  Melville]{mcinnes_umap_2020}
McInnes,~L.; Healy,~J.; Melville,~J. {UMAP}: {Uniform} {Manifold}
  {Approximation} and {Projection} for {Dimension} {Reduction}. 2020;
  \url{http://arxiv.org/abs/1802.03426}, arXiv:1802.03426 [stat]\relax
\mciteBstWouldAddEndPuncttrue
\mciteSetBstMidEndSepPunct{\mcitedefaultmidpunct}
{\mcitedefaultendpunct}{\mcitedefaultseppunct}\relax
\EndOfBibitem
\bibitem[Pedregosa \latin{et~al.}(2011)Pedregosa, Varoquaux, Gramfort, Michel,
  Thirion, Grisel, Blondel, Prettenhofer, Weiss, Dubourg, Vanderplas, Passos,
  Cournapeau, Brucher, Perrot, and Duchesnay]{pedregosa_scikit-learn_2011}
Pedregosa,~F.; Varoquaux,~G.; Gramfort,~A.; Michel,~V.; Thirion,~B.;
  Grisel,~O.; Blondel,~M.; Prettenhofer,~P.; Weiss,~R.; Dubourg,~V.
  \latin{et~al.}  Scikit-learn: {Machine} {Learning} in {Python}. \emph{J.
  Mach. Learn. Res.} \textbf{2011}, \emph{12}, 2825--2830\relax
\mciteBstWouldAddEndPuncttrue
\mciteSetBstMidEndSepPunct{\mcitedefaultmidpunct}
{\mcitedefaultendpunct}{\mcitedefaultseppunct}\relax
\EndOfBibitem
\bibitem[Kingma and Ba(2017)Kingma, and Ba]{kingma_adam_2017}
Kingma,~D.~P.; Ba,~J. Adam: {A} {Method} for {Stochastic} {Optimization}. 2017;
  \url{http://arxiv.org/abs/1412.6980}, arXiv:1412.6980 [cs]\relax
\mciteBstWouldAddEndPuncttrue
\mciteSetBstMidEndSepPunct{\mcitedefaultmidpunct}
{\mcitedefaultendpunct}{\mcitedefaultseppunct}\relax
\EndOfBibitem
\bibitem[Lundberg and Lee(2017)Lundberg, and Lee]{lundberg_unified_2017}
Lundberg,~S.; Lee,~S.-I. A {Unified} {Approach} to {Interpreting} {Model}
  {Predictions}. 2017; \url{https://arxiv.org/abs/1705.07874v2}\relax
\mciteBstWouldAddEndPuncttrue
\mciteSetBstMidEndSepPunct{\mcitedefaultmidpunct}
{\mcitedefaultendpunct}{\mcitedefaultseppunct}\relax
\EndOfBibitem
\bibitem[Rana \latin{et~al.}(2021)Rana, Brangwynne, and
  Panagiotopoulos]{rana_phase_2021}
Rana,~U.; Brangwynne,~C.~P.; Panagiotopoulos,~A.~Z. Phase separation vs
  aggregation behavior for model disordered proteins. \emph{The Journal of
  Chemical Physics} \textbf{2021}, \emph{155}, 125101\relax
\mciteBstWouldAddEndPuncttrue
\mciteSetBstMidEndSepPunct{\mcitedefaultmidpunct}
{\mcitedefaultendpunct}{\mcitedefaultseppunct}\relax
\EndOfBibitem
\bibitem[Das \latin{et~al.}(2018)Das, Eisen, Lin, and Chan]{das_lattice_2018}
Das,~S.; Eisen,~A.; Lin,~Y.-H.; Chan,~H.~S. A {Lattice} {Model} of
  {Charge}-{Pattern}-{Dependent} {Polyampholyte} {Phase} {Separation}.
  \emph{The Journal of Physical Chemistry B} \textbf{2018}, \emph{122},
  5418--5431, Publisher: American Chemical Society\relax
\mciteBstWouldAddEndPuncttrue
\mciteSetBstMidEndSepPunct{\mcitedefaultmidpunct}
{\mcitedefaultendpunct}{\mcitedefaultseppunct}\relax
\EndOfBibitem
\bibitem[Mao \latin{et~al.}(2010)Mao, Crick, Vitalis, Chicoine, and
  Pappu]{mao_net_2010}
Mao,~A.~H.; Crick,~S.~L.; Vitalis,~A.; Chicoine,~C.~L.; Pappu,~R.~V. Net charge
  per residue modulates conformational ensembles of intrinsically disordered
  proteins. \emph{Proceedings of the National Academy of Sciences of the United
  States of America} \textbf{2010}, \emph{107}, 8183--8188\relax
\mciteBstWouldAddEndPuncttrue
\mciteSetBstMidEndSepPunct{\mcitedefaultmidpunct}
{\mcitedefaultendpunct}{\mcitedefaultseppunct}\relax
\EndOfBibitem
\bibitem[Hand and Till(2001)Hand, and Till]{hand_simple_2001}
Hand,~D.~J.; Till,~R.~J. A {Simple} {Generalisation} of the {Area} {Under} the
  {ROC} {Curve} for {Multiple} {Class} {Classification} {Problems}.
  \emph{Machine Learning} \textbf{2001}, \emph{45}, 171--186\relax
\mciteBstWouldAddEndPuncttrue
\mciteSetBstMidEndSepPunct{\mcitedefaultmidpunct}
{\mcitedefaultendpunct}{\mcitedefaultseppunct}\relax
\EndOfBibitem
\bibitem[von B{\"u}low \latin{et~al.}(2025)von B{\"u}low, Tesei, and
  Lindorff-Larsen]{von_bulow_machine_2025}
von B{\"u}low,~S.; Tesei,~G.; Lindorff-Larsen,~K. Machine learning methods to
  study sequence--ensemble--function relationships in disordered proteins.
  \emph{Current Opinion in Structural Biology} \textbf{2025}, \emph{92},
  103028\relax
\mciteBstWouldAddEndPuncttrue
\mciteSetBstMidEndSepPunct{\mcitedefaultmidpunct}
{\mcitedefaultendpunct}{\mcitedefaultseppunct}\relax
\EndOfBibitem
\bibitem[Sidky \latin{et~al.}(2018)Sidky, Col{\'{o}}n, Helfferich, Sikora,
  Bezik, Chu, Giberti, Guo, Jiang, Lequieu, Li, Moller, Quevillon, Rahimi,
  Ramezani-Dakhel, Rathee, Reid, Sevgen, Thapar, Webb, Whitmer, and
  de~Pablo]{R:2018_SSAGES_JCP}
Sidky,~H.; Col{\'{o}}n,~Y.~J.; Helfferich,~J.; Sikora,~B.~J.; Bezik,~C.;
  Chu,~W.; Giberti,~F.; Guo,~A.~Z.; Jiang,~X.; Lequieu,~J. \latin{et~al.}
  {SSAGES}: Software Suite for Advanced General Ensemble Simulations. \emph{The
  Journal of Chemical Physics} \textbf{2018}, \emph{148}, 044104\relax
\mciteBstWouldAddEndPuncttrue
\mciteSetBstMidEndSepPunct{\mcitedefaultmidpunct}
{\mcitedefaultendpunct}{\mcitedefaultseppunct}\relax
\EndOfBibitem
\bibitem[McCarty \latin{et~al.}(2019)McCarty, Delaney, Danielsen, Fredrickson,
  and Shea]{mccarty_complete_2019}
McCarty,~J.; Delaney,~K.~T.; Danielsen,~S. P.~O.; Fredrickson,~G.~H.;
  Shea,~J.-E. Complete {Phase} {Diagram} for {Liquid}--{Liquid} {Phase}
  {Separation} of {Intrinsically} {Disordered} {Proteins}. \emph{The Journal of
  Physical Chemistry Letters} \textbf{2019}, \emph{10}, 1644--1652, Publisher:
  American Chemical Society\relax
\mciteBstWouldAddEndPuncttrue
\mciteSetBstMidEndSepPunct{\mcitedefaultmidpunct}
{\mcitedefaultendpunct}{\mcitedefaultseppunct}\relax
\EndOfBibitem
\bibitem[Saar \latin{et~al.}(2023)Saar, Qian, Good, Morgunov,
  Collepardo-Guevara, Best, and Knowles]{saar_theoretical_2023}
Saar,~K.~L.; Qian,~D.; Good,~L.~L.; Morgunov,~A.~S.; Collepardo-Guevara,~R.;
  Best,~R.~B.; Knowles,~T. P.~J. Theoretical and {Data}-{Driven} {Approaches}
  for {Biomolecular} {Condensates}. \emph{Chemical Reviews} \textbf{2023},
  \emph{123}, 8988--9009, Publisher: American Chemical Society\relax
\mciteBstWouldAddEndPuncttrue
\mciteSetBstMidEndSepPunct{\mcitedefaultmidpunct}
{\mcitedefaultendpunct}{\mcitedefaultseppunct}\relax
\EndOfBibitem
\end{mcitethebibliography}
\end{document}


\title{Supplementary Information \\
for\\
When $B_2$ is Not Enough: Evaluating Simple Metrics for Predicting Phase Separation of Intrinsically Disordered Proteins}

\author{Wesley W. Oliver$^1$, William M. Jacob$^2$, and Michael A. Webb$^{1*}$\\
\\
{\small $^1$Department of Chemical and Biological Engineering, Princeton University, Princeton, NJ 08544, USA}\\
{\small $^2$Department of Chemistry, Princeton University, Princeton, NJ 08544, USA}\\
{\small $^*$Corresponding Author: mawebb@princeton.edu}}

\date{}
\maketitle

\tableofcontents
\newpage

\section{Example Equations-of-State\label{sec:si_eos}}
To circumvent the high cost of direct-coexistence simulations, we instead approximate the phase behavior of a sequence by searching for a negative pressure loop in its equation of state (EoS) (``Assessment of Phase Behavior'' in Section 2.2 of Methods).
Example EoSs are shown for a phase-separating and non-phase-separating sequence in Figure \ref{fig:eos} below.
For the phase-separating case, the approximation of the condensed phase density $\rho_\mathrm{c}$---the largest value of $\rho$ satisfying $P\left(\rho\right)=0$---is also shown.

\begin{figure}[ht]
    \centering
    \includegraphics{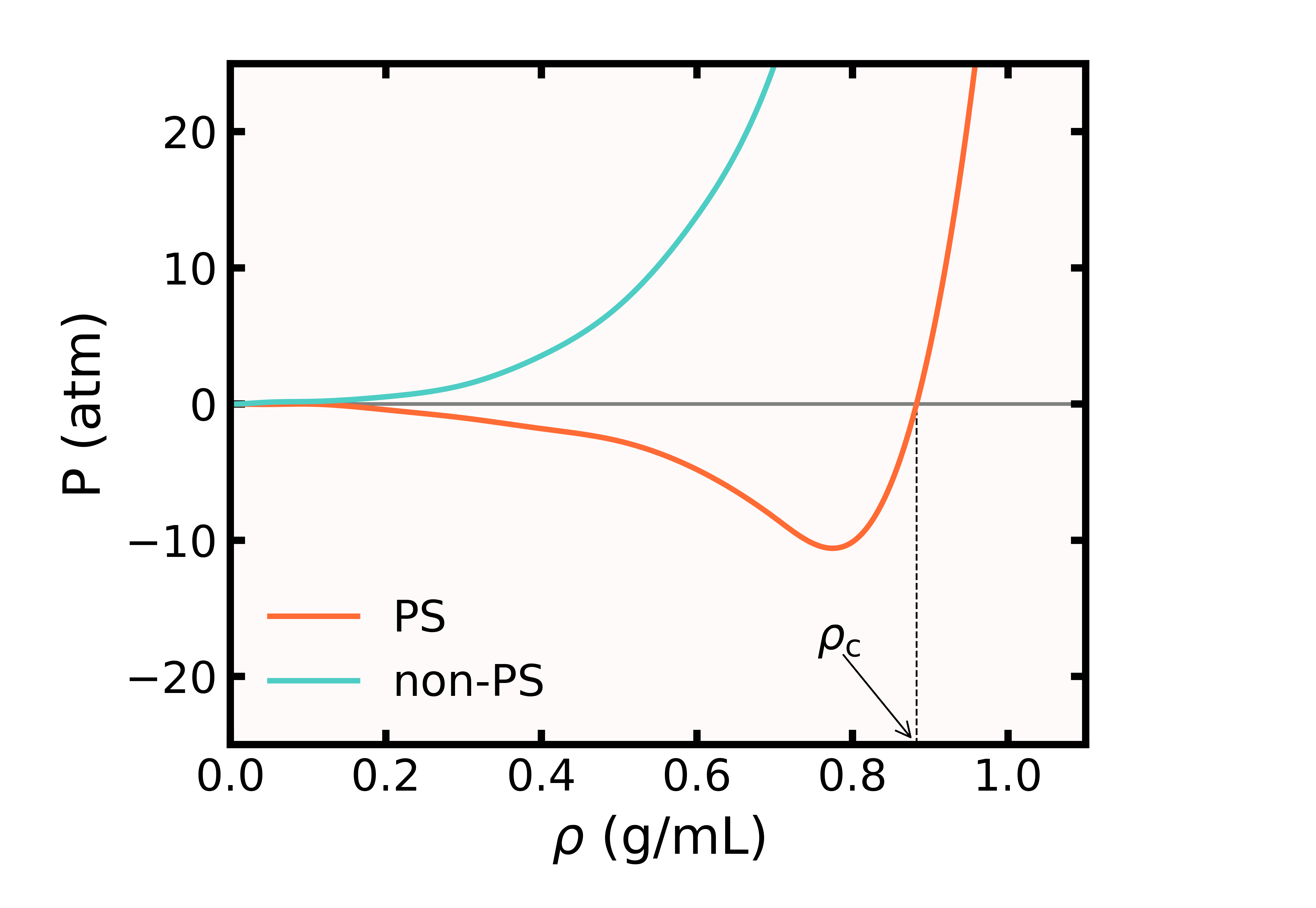}
    \caption{Representative equations-of-state (EoS) for a phase-separating (PS) and non-phase-separating (non-PS) IDP sequence. EoSs are constructed by generating a cubic spline between $\left(\rho,P\right)$ values (``Assessment of Phase Behavior'' in Section 2.2 of Methods). If negative pressure values are observed at any density, a sequence is presumed to undergo phase separation in the thermodynamic limit, and the largest zero of the constructed cubic spline is taken to be the condensed-phase density, $\rho_\mathrm{c}$.}
    \label{fig:eos}
\end{figure}

\newpage
\section{Effect of Normalization on Metric Classification Accuracy}
The second virial coefficient $B_2$ and radius of gyration $R_\mathrm{g}$ of a polymer are extensive quantities that scale with length.
To create normalized, intensive versions of each metric, we define $\tilde{B}_2=B_2/V_0$ where $V_0=\left(4\pi/3\right)b_0^3\left(N/6\right)^\frac{3}{2}$ and $\tilde{R}_\mathrm{g} = R_\mathrm{g} / R_\text{g}^{\text{id}}$ where $R_\text{g}^{\text{id}} \equiv \sqrt{\langle R_\text{g}^2 \rangle_\text{id}} = b_0\sqrt{N/6}$ given a polymer of length $N$ and bond length $b_0$~\cite{rubinstein_polymer_2003}.
We set $b_0=3.82$ \unit{\angstrom} to reflect the equilibrium bond length of the HPS-Urry model. 
Figure \ref{fig:metric_normaliztion} below shows that these normalized quantities provide a better predictor of phase behavior as measured by the AUC of a threshold classifier than their non-normalized counterparts.

\begin{figure}[ht]
    \centering
    \includegraphics{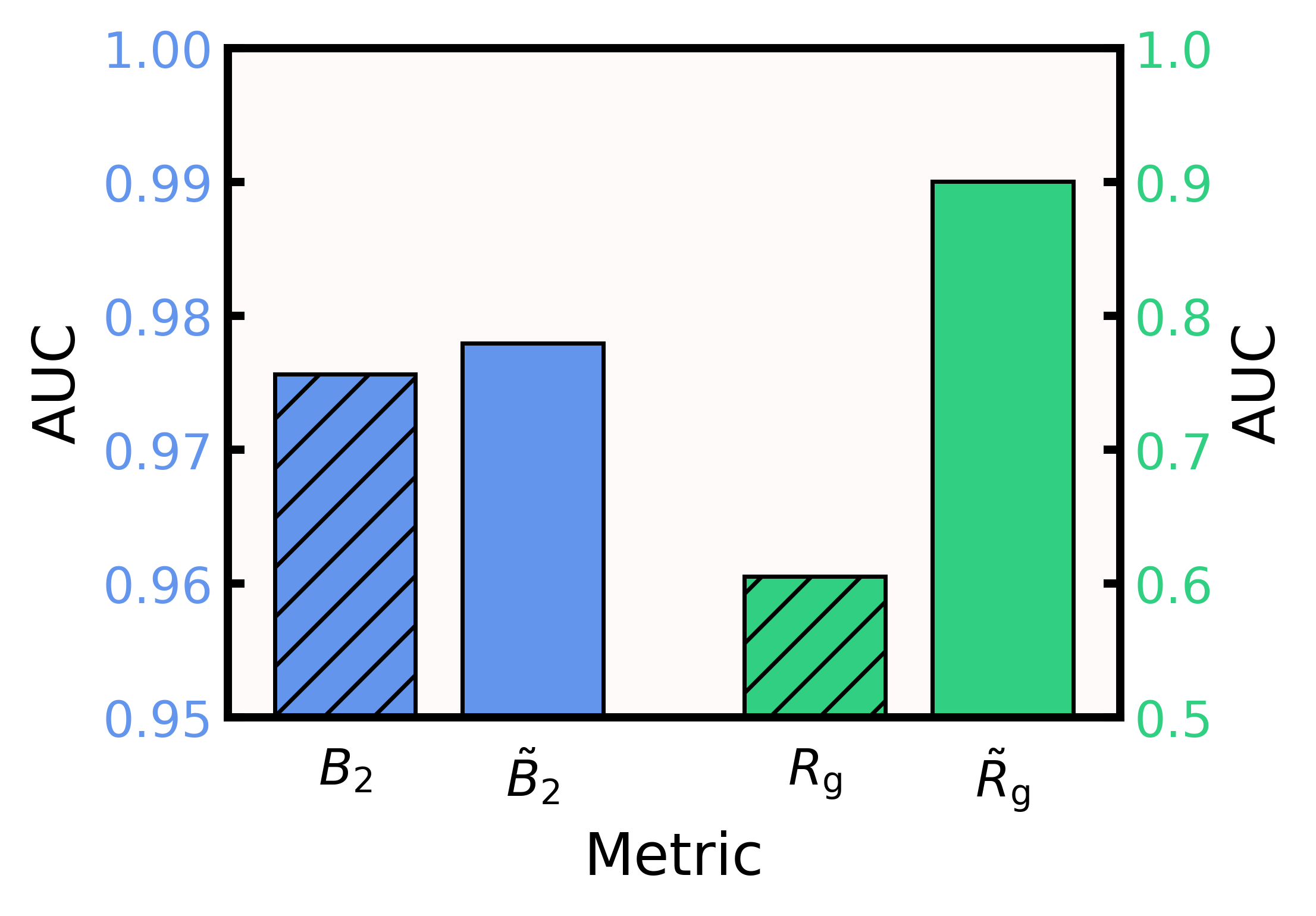}
    \caption{Impact of normalization on the classification accuracy of $B_\mathrm{2}$ and $R_\mathrm{g}$. Classification accuracy is assessed by AUC. We define $\tilde{B}_2=B_2/V_0$ and $\tilde{R}_\mathrm{g}=R_\mathrm{g}/\langle R_\mathrm{g}\rangle_\mathrm{ideal}$ to account for the extensive relationships of $B_\mathrm{2}$ and $R_\mathrm{g}$ with sequence length. The factor $V_0$ stems from the pervaded volume of an ideal chain, and $\langle R_\mathrm{g}\rangle_\mathrm{ideal}$ stems from the radius of gyration of an ideal chain. The hashed and solid boxes respectively represent the non-normalized and normalized metrics.}
    \label{fig:metric_normaliztion}
\end{figure}

\newpage
\section{Expenditure Density Sensitivity Analysis\label{sec:si_sensitivity_analysis}}
The expenditure density $\rho^\mathrm{*}$ is a newly proposed metric of propensity for phase separation that captures the density to which a sequence
can be reversibly compressed from infinite dilution given an allowance of work.
Given an equation of state (EoS) $P\left(\rho\right)$, $\rho^\mathrm{*}$ is calculated via
\begin{equation}\label{eq:exp}
    \hat{w}^*=\frac{W^*}{m}=\int_0^{\rho^\mathrm{*}} d\rho\,\frac{P\left(\rho\right)}{\rho^2}
\end{equation}
where $\hat{w}^\mathrm{*}$ is an intensive work allowance that must be chosen to solve the equality.

Figure \ref{fig:sensitivity_analysis} below shows the sensitivity of the classification accuracy of $\rho^\mathrm{*}$ as assessed by the AUC of a threshold classifier to choices of $\hat{w}^\mathrm{*}$.
The accuracy of $\rho^\mathrm{*}$ peaks at approximately $\hat{w}^*=15$ \unit{atm\cdot\milli\liter\per\gram}.
Below this value, accuracy drops off rapidly due to insufficient integration of the EoS, while accuracy is relatively robust to further increases in $\hat{w}^*$.

\begin{figure}[ht]
    \centering
    \includegraphics{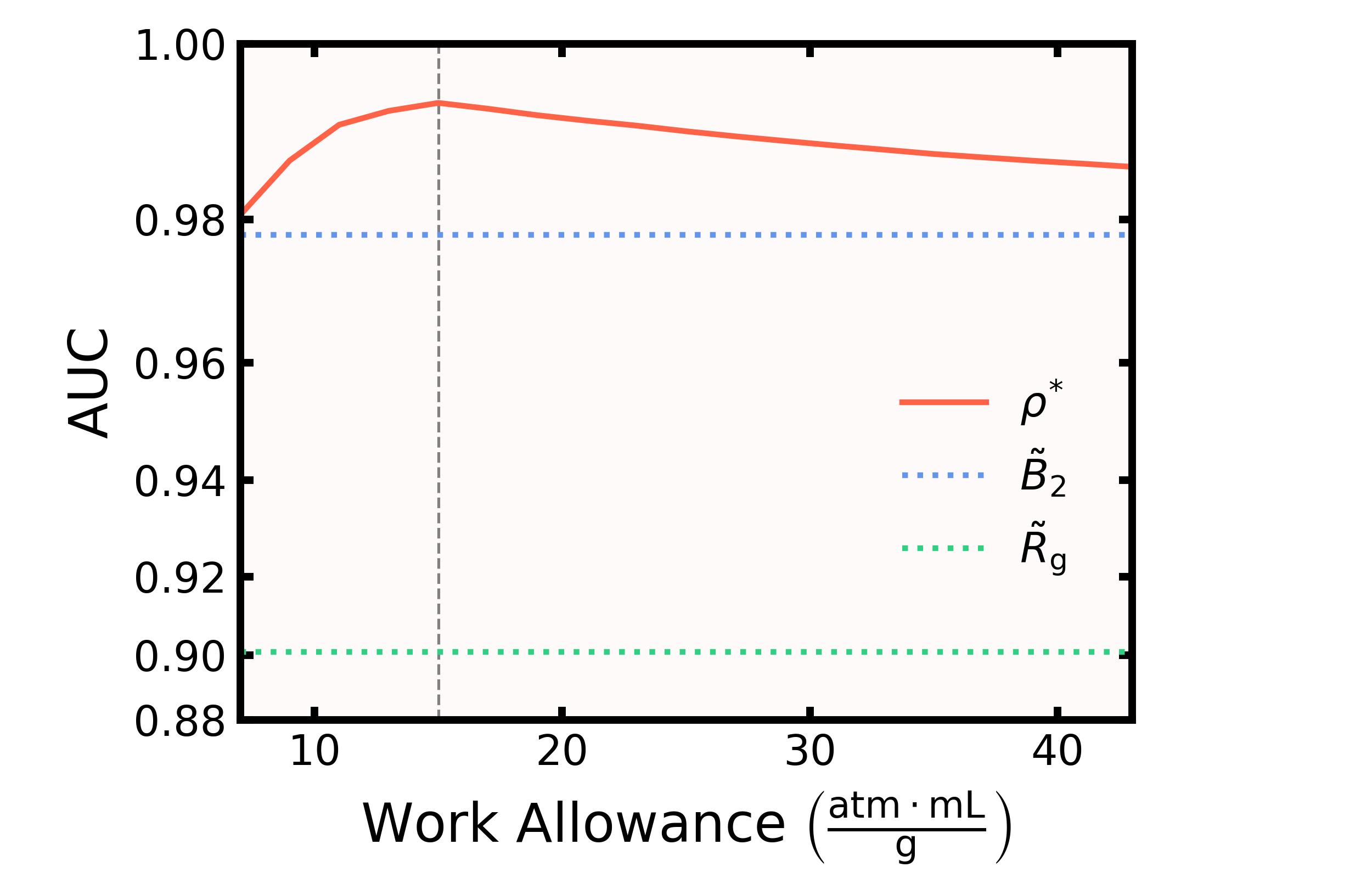}
    \caption{Sensitivity analysis of the classification accuracy of $\rho^\mathrm{*}$. Accuracy is assessed by AUC and shown for different choices of the intensive work allowance, $\hat{w}^*$. A maximum AUC is observed at a $\hat{w}^*$ of approximately 15 \unit{atm\cdot\milli\liter\per\gram} (vertical dashed line). For reference, the AUC values of $\tilde{B}_2$ and $\tilde{R}_\mathrm{g}$ are also plotted.}
    \label{fig:sensitivity_analysis}
\end{figure}


\newpage
\section{Sequence Featurization\label{sec:si_sequence_featurization}}
For each sequence, we construct an augmented scaled fingerprint consisting of 30 features~\cite{patel_data-driven_2023}. These features and their formulae are listed below.

\begin{itemize}
\item the composition of each amino acid in the sequence, $f_X$, where $\mathrm{X}$ corresponds to one of the amino acids (A, C, D, E, F, G, H, I, K, L, M, N, P, Q, R, S, T, V, W, and Y) and $\sum_X f_X = 1$;

\item the sequence length (i.e., number of amino acids), $N$

\item the fraction of positively charged residues, $f_+$

\item the fraction of negatively charged residues, $f_-$

\item the absolute value of the net charge per residue, $\overline{|q_i|}$,
\begin{equation}
\overline{|q_i|} = \frac{1}{N}\left|\sum_{i=1}^N q_i\right|
\end{equation}

\item the sequence charge decoration, SCD,
\begin{equation}
\text{SCD} = \frac{1}{N}\sum_{i=1}^N \sum_{j=i+1}^N q_i q_j (j-i)^{1/2}
\end{equation}

\item the sequence hydropathy decoration, SHD,
\begin{equation}
\text{SHD} = \frac{1}{N}\sum_{i=1}^N \sum_{j=i+1}^N (\lambda_i + \lambda_j)(j-i)^{-1}
\end{equation}

\item the average hydrophobicity per residue, $\overline{\lambda_i}$,
\begin{equation}
\overline{\lambda_i} = \frac{1}{N}\sum_{i=1}^N \lambda_i
\end{equation}
where $\lambda_i$ are hydrophobicity values of residue $i$ derived from the Urry hydrophobicity scale~\cite{urry_hydrophobicity_1992}.

\item the Shannon entropy, $S$,
\begin{equation}
S = -\sum_X f_X \log f_X
\end{equation}

\item the average molecular weight, $m_i$,
\begin{equation}
\overline{m_i} = \frac{1}{N}\sum_{i=1}^N m_i
\end{equation}

where $m_i$ is the molecular weight of residue $i$ in the sequence

\item a mean-field prediction of the second-virial coefficient, $B_2^\mathrm{MF}$,
\begin{equation}
B_2^\mathrm{MF}= \sum_{i=1}^N \sum_{j=1}^N b_{2,ij}
\end{equation}

where $b_{2,ij} = 2\pi \int_0^\infty dr \, r^2 \left[1 - e^{-\beta u_{ij}(r)}\right]$ is the second-virial coefficient between monomers $i$ and $j$, which are assumed to be on different chains, and $u_{ij}(r)$ is the pair potential between monomers $i$ and $j$

\end{itemize}

\newpage
\section{Hyperparameter Grid\label{sec:si_hyperparameter_grid}}
The hyperparameters of all neural network-based models (Section 2.4 of Methods) are tuned using grid-search cross-validation.
Hyperparameters are tuned over the following ranges:

\begin{itemize}
    \item Hidden layers: \{(10), (100), (100,100), (10,10), (50,50)\}
    \item Activation: \{tanh, ReLU, logistic\}
    \item Batch size: \{8, 16, 32, 64\}$^*$
    \item Learning rate: \{0.0001, 0.001, 0.01\}
    \item L2 regularization ($\alpha$): \{0.0001, 0.001, 0.01\}
\end{itemize}
$^*$Batch sizes are filtered to ensure they are less than 70\% of the training set size. For training sets with $\leq$30 samples, batch size 4 was added to the grid.